\documentclass[final,5p,times]{elsarticle}
\usepackage{graphicx}
\usepackage{bm}
\usepackage{booktabs}
\usepackage{amssymb}
\usepackage{multirow}
\usepackage{amsmath}
\usepackage{dcolumn}
\usepackage{hyperref}
\usepackage{color,units}
\usepackage[dvipsnames]{xcolor}
\usepackage{lineno}
\usepackage{xspace}
\usepackage[normalem]{ulem}
\usepackage{float}
\usepackage{orcidlink}
\usepackage{tabularray}
\usepackage{placeins}
\definecolor{DarkGreen}{rgb}{0.0, 0.6, 0.3}


\usepackage[T1]{fontenc}
\usepackage[english]{babel}
\usepackage{microtype}

\NewDocumentCommand{\codeword}{v}{%
\texttt{\textcolor{blue}{#1}}%
}

\begin{document}
\begin{frontmatter}

\title{Enhancing Event Reconstruction in Hyper-Kamiokande with Machine Learning: \\A ResNet Implementation}

\affiliation[monash]{
  organization={School of Physics and Astronomy, Monash University},
  city={Clayton},
  state={VIC},
  postcode={3800},
  country={Australia}
}

\affiliation[ozgrav]{
  organization={OzGrav: The ARC Centre of Excellence for Gravitational Wave Discovery},
  city={Clayton},
  state={VIC},
  postcode={3800},
  country={Australia}
}

\affiliation[imperial]{
  organization={Imperial College London, Department of Physics},
  city={London},
  country={United Kingdom}
}

\affiliation[icrr]{
  organization={Research Center for Cosmic Neutrinos, Institute for Cosmic Ray Research, The University of Tokyo},
  city={Kashiwa},
  state={Chiba},
  postcode={277-8582},
  country={Japan}
}

\affiliation[kavli]{
  organization={Kavli Institute for the Physics and Mathematics of the Universe (WPI), The University of Tokyo Institutes for Advanced Study, University of Tokyo},
  city={Kashiwa},
  state={Chiba},
  postcode={277-8583},
  country={Japan}
}

\affiliation[TRIUMF]{
  organization={TRIUMF},
  addressline={4004 Wesbrook Mall},
  city={Vancouver},
  state={BC},
  postcode={V6T 2A3},
  country={Canada}
}

\affiliation[unimelb]{
  organization={School of Physics, The University of Melbourne},
  city={Parkville},
  state={VIC},
  postcode={3010},
  country={Australia}
}

\author[monash]{Andrew Atta}

\author[imperial]{Nick Prouse}

\author[icrr]{Shuoyu Chen}

\author[TRIUMF]{Felix Cormier}

\author[TRIUMF]{Wojciech Fedorko}

\author[icrr,kavli]{Kimihiro Okumura}

\author[kavli]{Patrick de Perio}

\author[monash,ozgrav]{Eric Thrane}

\author[unimelb]{Phillip Urquijo}

\begin{abstract}
The forthcoming Hyper-Kamiokande experiment is expected to achieve unprecedented sensitivity to neutrino oscillations. Meeting these precision goals will require substantially larger Monte Carlo datasets than previous experiments had to satisfy the experiment’s stringent systematic-uncertainty requirements. While traditional maximum-likelihood reconstruction algorithms provide high-quality event reconstruction, their per-event computational cost makes the reconstruction of these large, systematically varied Monte Carlo samples increasingly impractical.
In this work, we demonstrate a neural-network-based reconstruction approach for the Hyper-Kamiokande far detector using simulated data with the experiment’s detector geometry.
Single-particle events are generated with kinetic energies ranging from the Cherenkov threshold up to \unit[2]{GeV} above threshold and propagated through the detector. The resulting photomultiplier tube charge and timing information are mapped to $274\times144$ five-channel images that serve as inputs to ResNet models implemented in the \textsc{WatChMaL} framework. These models are trained to (i) classify events into four particle hypotheses ($e$, $\mu$, $\gamma$, $\pi^{0}$) and (ii) regress the interaction vertex, direction, and momentum of electrons and muons.
Averaged over the full kinematic range, the regression models achieve momentum resolutions of $1.27\%$ for muons and $2.21\%$ for electrons, angular resolutions of $0.91^\circ$ and $1.60^\circ$, and vertex resolutions of $13.6\,\mathrm{cm}$ and $17.0\,\mathrm{cm}$, respectively, that are better than those reported for traditional reconstruction methods in similar studies. The classifier improves $e$-$\mu$, $e$-$\gamma$, and $e$-$\pi^{0}$ separation relative to the traditional reconstruction baseline, with corresponding areas under the ROC curve of $0.9999995$, $0.642$, and $0.9659$.
Crucially, our networks achieve inference times of 1-2~ms per event on a single GPU, yielding speed-ups of $3.2\times10^{4}$-$5.2\times10^{4}$ relative to likelihood-based reconstruction, and highlighting deep learning as a scalable alternative for Hyper-Kamiokande event reconstruction.
\end{abstract}

\end{frontmatter}

\section{Introduction}\label{sec:intro}

Hyper-Kamiokande~\cite{Abe2018, HK2015, HK2025} is a next-generation water Cherenkov experiment currently under construction in Japan. Building on the success of Kamiokande~\cite{Koshiba1984Kamiokande}, Super-Kamiokande~\cite{SK1998,fukuda2003}, and the long-baseline T2K program~\cite{T2K2011,T2K2023}, Hyper-Kamiokande is designed to deliver precision measurements of neutrino oscillation parameters, with particular sensitivity to the CP-violating phase $\delta_{\mathrm{CP}}$. These measurements will be enabled by a planned $\unit[1.3]{MW}$ $\nu_{\mu}$ beam produced at the J-PARC accelerator and directed toward the detector over a $\unit[295]{km}$ baseline~\cite{Abe2019}.

In water Cherenkov detectors such as Hyper-Kamiokande, neutrino-nucleus interactions produce charged secondary particles that emit Cherenkov radiation as they traverse the water at velocities exceeding the phase velocity of light in the medium~\cite{Abe2018, Cherenkov1937}. This light is detected by photomultiplier tubes (PMTs) lining the detector walls, which record photo-electron charges (the electric charge produced when a photon liberates an electron from the PMT photocathode) and hit times that typically form ring-like patterns on the cylindrical detector surface.

The PMT charge and timing information constitute the primary observables used to reconstruct the interaction vertex, the directions and momenta of the charged particle tracks, and the visible energy~\cite{Lalakulich2012, HK2015}. The visible energy is the energy inferred from the observed Cherenkov light yield and is generally less than the true total energy of the interaction. The dominant invisible contributions come from neutrons, from hadrons below the Cherenkov threshold (in particular recoil protons, whose threshold lies near $\unit[485]{MeV}$ of kinetic energy in water), from low-energy de-excitation photons that produce only sub-threshold electrons, and from nuclear removal energy \citep{Dolan2026}. The momenta of individual tracks are estimated from the total charge collected, since the amount of Cherenkov light produced is related to the particle's energy loss along its track. These reconstructed quantities enable an estimate of the incident neutrino energy $E_{\nu}$, a key input to measurements of the appearance and disappearance probabilities~\cite{T2K2023, HK2025}.

The primary component of the experiment, the Hyper-Kamiokande far detector, will consist of a $\unit[258]{kton}$ ($\unit[188]{kton}$ fiducial) water Cherenkov detector instrumented with approximately $20{,}000$ inward-facing $\unit[50]{cm}$ PMTs, $\sim800$ multi-PMT (mPMT) modules, and 3,600 3-inch outer-detector PMTs~\cite{HK2025, Kose2025}. Each mPMT module comprises a compact assembly of 19 3-inch PMTs housed within a single pressure vessel, providing enhanced directional sensitivity, improved photon counting, and increased redundancy relative to single large-area PMTs. The outer-detector PMTs form an independent veto layer surrounding the inner volume, enabling the identification of entering cosmic rays and escaping particles.

Determining the CP-violating phase $\delta_{\mathrm{CP}}$ requires precise measurements of the $\nu_{e}$ appearance probability $P(\nu_{\mu}\rightarrow\nu_{e})$~\cite{T2K2023}. Since neutrino oscillation probabilities depend on neutrino energy $E_{\nu}$~\cite{PDG2024,Bilenky1978}, accurate reconstruction of the energy and direction of the outgoing charged lepton in charged-current interactions is essential. Precise determination of the interaction vertex is also required, as it defines whether an event lies within the detector’s fiducial volume and directly impacts the measured event rates.

In water Cherenkov detectors, charged particles produce characteristic ring topologies that enable particle identification. Relativistic muons form sharp, well-defined rings corresponding to minimum-ionizing tracks, whereas electrons initiate electromagnetic showers that yield broader, fuzzier rings with less distinct edges. These morphological differences underpin the traditional separation of “$\mu$-like’’ and “$e$-like’’ events. More challenging backgrounds arise from neutral-current $\pi^{0}$ production, where the $\pi^{0}\rightarrow\gamma\gamma$ decay produces two electromagnetic showers. At sufficiently high momenta, the two rings overlap and mimic a single $e$-like ring, constituting the second most dominant background to $\nu_{\mu}\rightarrow\nu_{e}$ appearance and thus directly impacting measurements of the CP-violating phase $\delta_{\mathrm{CP}}$. Separating electrons from single photons is similarly difficult: although photons do not emit Cherenkov light before converting, the resulting electromagnetic shower produces an $e$-like topology, especially when the conversion point lies near the primary vertex. This task is, in fact, so difficult that an $e$-$\gamma$ separator has not yet been successfully achieved in Super-Kamiokande. Robust $e$-$\pi^{0}$ and $e$-$\gamma$ discrimination is therefore essential for maintaining high-purity $\nu_{e}$ samples and suppressing neutral-current backgrounds in oscillation analyses.

Precision oscillation analyses rely on large Monte Carlo samples and ensembles of pseudo-experiments in which flux, cross-section, and detector-response systematics are propagated to reconstructed observables. In this forward-folded, template-based framework, high-statistics simulated samples are used to construct predicted event distributions, efficiencies, and migration matrices, while oscillation and many systematic parameters are applied through reweighting. The unprecedented statistics of Hyper-Kamiokande sharpen this requirement: controlling Monte Carlo statistical fluctuations well below the percent-level systematic budget demands correspondingly larger reconstructed samples than previous experiments required.

Crucially, the cost of this program is not dominated by the oscillation analysis itself. While oscillation parameter scans are computationally inexpensive, and many systematic parameters can be applied by reweighting, the detector and interaction systematics that alter the reconstructed observables, such as detector-calibration shifts or alternative final-state interaction models, each require a separately reconstructed Monte Carlo sample. These reconstructed datasets underpin the estimation of selection efficiencies, energy-scale uncertainties, and event migration across fiducial boundaries, as well as the pseudo-experiment ensembles used to assess sensitivity to CP violation, mass ordering, and rare backgrounds such as neutral-current $\pi^{0}$ production. The run time of producing each such sample is dominated by the reconstruction step, and the current state-of-the-art likelihood-based reconstruction \textsc{fiTQun} (see~\ref{sec:fitqun}) carries an $\mathcal{O}(\mathrm{min})$ per-event cost~\cite{Prouse2023}. Reconstructing the full suite of systematically varied samples with \textsc{fiTQun} is therefore prohibitive: the reconstruction stage, not the analysis, sets the practical limit on how many systematic variations can be considered. This challenge is compounded by the fact that likelihood-based techniques such as \textsc{fiTQun} often struggle with complex event topologies, including overlapping Cherenkov rings from photon conversions, vertices near the detector wall, or events with sparse photon statistics. Meeting Hyper-Kamiokande's precision goals will therefore require a reconstruction that is scalable and fast while remaining at least as performant as \textsc{fiTQun}.

Machine-learning based reconstruction offers a promising path toward meeting these computational and topological challenges. Convolutional neural networks \cite{krizhevsky2012}, and in particular residual networks (ResNets) \cite{He2016}, are well suited to process the high-dimensional charge and time images recorded by the PMT array. At Super-Kamiokande, machine-learning techniques are already employed for solar-neutrino classification \cite{alejandro2023} and for neutron-capture tagging \cite{Jamieson2022}, and they show encouraging results in neutrino-reconstruction studies for other experiments \cite{Aguilar2025, Qian2021, Aurisano2016, Mondal2024}. Within this landscape, the Water Cherenkov Machine Learning (\textsc{WatChMaL}) framework \cite{watchmal2025} has been developed. Prouse et al.~\cite{Prouse2023} demonstrated that a ResNet-50 model trained on simulations of the Intermediate Water Cherenkov Detector (another component of the Hyper-Kamiokande experimental configuration) matches or surpasses \textsc{fiTQun} resolutions while reducing inference time by several orders of magnitude. Yet to date, no published study has applied these deep-learning methods to the Hyper-Kamiokande far detector.

In this paper, we present a demonstration of machine-learning-based event reconstruction for the Hyper-Kamiokande far detector using simulated data. We develop models for four-class particle identification ($e$, $\mu$, $\gamma$, $\pi^{0}$) and for reconstruction of the interaction vertex, direction, and lepton momentum for electrons and muons from Cherenkov threshold up to 2~GeV above threshold. The resulting reconstruction performance is evaluated using standard metrics and contextualized with respect to published results from traditional likelihood-based methods. We emphasize the implications of this approach for future $\nu_{\mu}\rightarrow\nu_{e}$ appearance measurements and $\delta_{\mathrm{CP}}$ sensitivity studies, particularly in the context of large-scale Monte Carlo production.

The remainder of this paper is organized as follows. Section~\ref{sec:simdata} describes the simulation and reconstruction methodology. Reconstruction performance for kinematic regression and particle classification is presented in Section~\ref{sec:results}. Section~\ref{sec:discussion} discusses the implications of these results, including reconstruction robustness and computational performance. Finally, Section~\ref{sec:conclusion} summarizes the main findings and outlines directions for future work.

Additional material is provided in the appendices.~\ref{sec:physics-background} provides physics background and motivation,~\ref{sec:fitqun} describes the traditional likelihood-based reconstruction used for benchmarking, and~\ref{sec:extended-diagnostics} presents supplementary reconstruction diagnostics.~\ref{sec:decayeffect} discusses the impact of muon decay on reconstruction performance,~\ref{sec:training-curves} presents representative training curves, and~\ref{sec:wcsim-config} documents the simulation configuration used to generate the datasets. ~\ref{app:double-cover} and~\ref{app:hit-time-invariance} describe additional data transformations prior to input into the networks. 

\section{Method} \label{sec:simdata}

\subsection{Simulated Data} \label{sec:data}

The models in this work take as input two-dimensional "images" of size $274\times144$ with five channels per pixel. Each image represents a single simulated event, in which a charged particle propagates through the water and radiates Cherenkov photons that are collected by the PMTs lining the detector wall. Events are generated with the Geant4-based open-source water-Cherenkov simulation package \textsc{WCSim}~\cite{WCSim}, which propagates the particles and the resulting optical photons through the detector geometry and digitizes the PMT response. The full simulation configuration, including detector geometry, optical properties, digitization and trigger settings, dark-noise model, and particle-generation commands, is provided in~\ref{sec:wcsim-config}; here we describe only the choices that directly impact the construction of the neural-network inputs.

\begin{figure}[ht]
 \centering
 \includegraphics[width=0.8\linewidth]{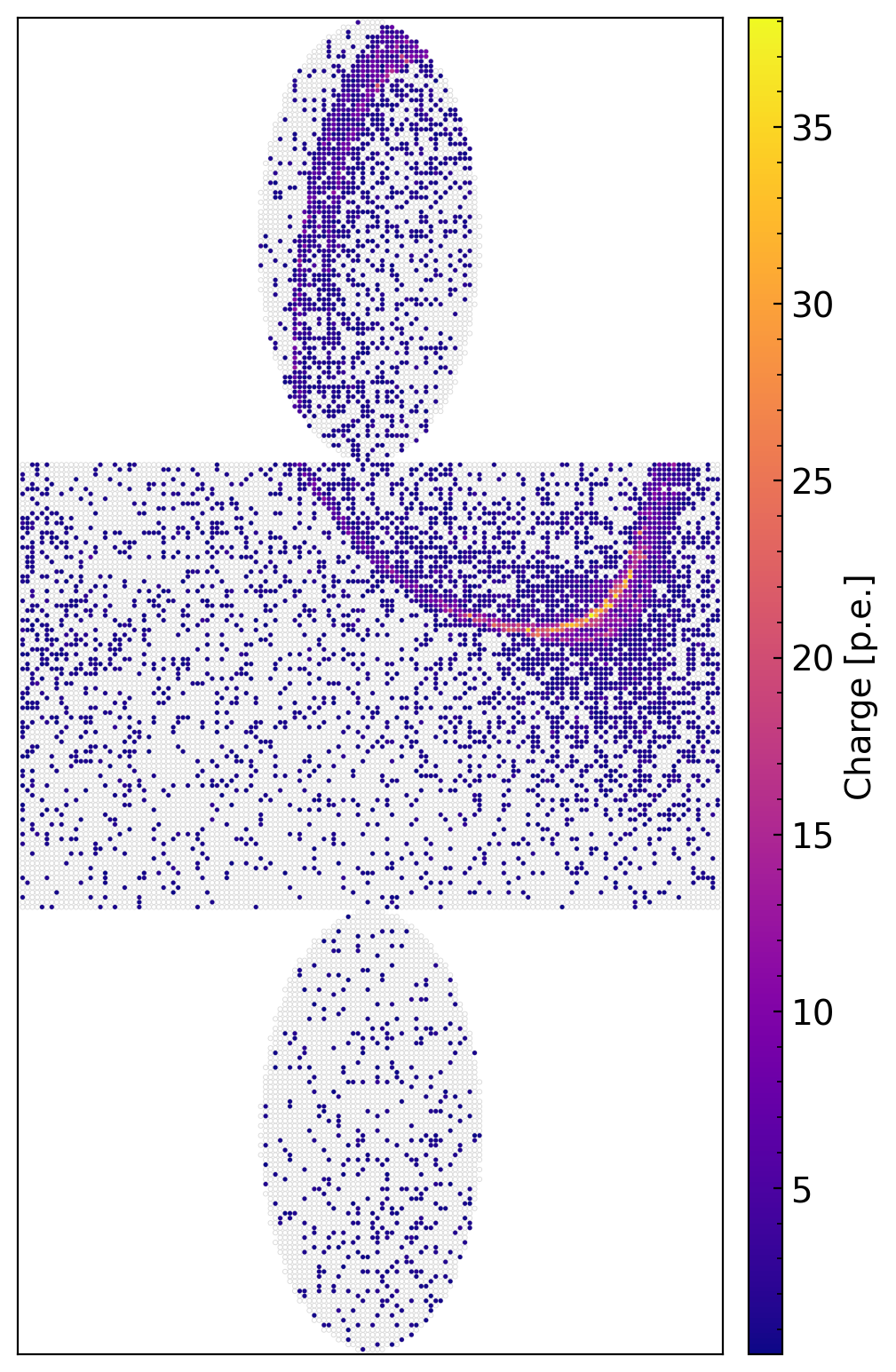}
 \caption{Example $274\times144$ image of a $\mu$ event. Only the integrated-charge channel is shown, in photo-electrons and prior to scaling; the remaining four channels store the PMT hit time and the three Cartesian coordinates of each PMT.}
 \label{fig:input-image}
\end{figure}

The image is obtained by unwrapping the cylindrical detector surface along the azimuthal direction, with the top and bottom endcaps placed perpendicular to the unraveled barrel. An example input image produced by this procedure, showing only the charge channel, is provided in Figure~\ref{fig:input-image}. This flattened representation largely preserves the adjacency of neighboring PMTs in the image, allowing convolutional kernels to exploit local correlations.

The PMTs of the actual detector are not, however, arranged on a rectangular grid on the detector wall. Instead they alternate between PMT and uninstrumented empty space. The spaces in each row are staggered such that there is one empty slot above, below, and to either side of each slot instrumented with a PMT. This spacing is what gives the inner detector its $20\%$ photocoverage~\citep{HK2025}, compared with $40\%$ in Super-Kamiokande. The mapping accommodates this by assigning two consecutive columns of PMTs to one column of pixels in the image, so that the PMTs are packed contiguously and no pixels are left empty between them. A conventional square kernel applied to this image therefore acts over a block of PMTs on the detector wall that physically represents a rectangle with width double its height.

To reduce the number of empty pixels and provide additional locations in the image where the barrel and endcaps connect physically nearby PMTs, the image is rearranged into a "double-cover" representation of the detector surface, following the implementation used for the Intermediate Water Cherenkov Detector in ref.~\citep{Prouse2023}. Details of exactly how this transformation is done can be found in \ref{app:double-cover}.

Each image has five channels. The first channel stores the PMT hit time, as reported by the \textsc{WCSim} digitizer, divided by $\unit[100]{ns}$. The readout window is opened when the number of PMTs registering hits within a 200~ns interval exceeds a threshold (nominally 25), defining the trigger time $T_0$. The window begins 400~ns before this trigger ($T_0 - 400~\text{ns}$) and ends 950~ns after ($T_0 + 950~\text{ns}$), giving a total width of $1.35\,\mu\mathrm{s}$.

The time assigned to a PMT is the leading edge of the first pulse it records within the window. The second channel stores all the charge of that pulse (i.e. the summed charge of the photo-electrons arriving within 200~ns of its leading edge) in units of photo-electrons. The remaining three channels store the Cartesian coordinates $(x,y,z)$ of the PMT corresponding to each pixel. Each channel is scaled so that its typical values are of order unity. This is done by dividing the hit times by 100, the hit charges by 10, and the three PMT coordinate channels by $\unit[10^{4}]{cm}$ each. PMTs that record no hit, and pixels that have no associated PMT, are assigned zero charge, while their entry in the time channel is set to the median of the hit times recorded in that event and zero in the position channels.

Although the full \textsc{WCSim} geometry includes the inner-detector multi-PMT modules and the outer veto region, only the 50\,cm PMTs of the inner detector are used for constructing the input images. The outer-detector PMTs are excluded from reconstruction, as they are used only during pre-processing and not in the reconstruction itself. The multi-PMTs are omitted from this study as their total photocathode coverage is small compared to that of the $\unit[50]{cm}$ PMTs --- each mPMT contains 19 three-inch PMTs, giving a total photocoverage of up to $0.4\%$. However, efforts are underway to include mPMT information in the input image which we leave for a future study.

We generate four statistically independent datasets (produced in separate simulation campaigns) each containing approximately $10^7$ events: single electrons ($e$), muons ($\mu$), $e^{-}e^{+}$ pair-producing gamma rays ($\gamma$), and neutral pions ($\pi^{0}$).
For each event, the interaction vertex is drawn uniformly throughout the inner-detector volume, the direction is sampled isotropically, and the true kinetic energy is chosen uniformly above Cherenkov threshold up to 2~GeV (exact numerical bounds are given in \ref{sec:wcsim-config}).

Muon decay is disabled in the muon sample to prevent Cherenkov light from decay electrons from being conflated with the primary muon signal.
This allows the networks to be trained on the isolated topology of a single muon, rather than on an event which contains Cherenkov light from both the initial muon and subsequent decay electron.
In a full reconstruction pipeline, hit-time clustering at the pre-processing stage will generally separate the prompt muon light from the delayed Michel-electron contribution, making it unnecessary for the networks to learn both components simultaneously in most cases. 
Nevertheless, a small fraction of events will contain Michel electrons within the readout gate that cannot be cleanly isolated through timing information alone. We show how this affects reconstruction in~\ref{sec:decayeffect}. Future extensions of this work, and the development of a fully neural-network-based reconstruction pipeline, will therefore need to accommodate composite event topologies to robustly handle these outliers.

All gamma-ray events are forced to undergo pair production. 
This is because non-converting gammas predominantly undergo Compton scattering, producing electron tracks that are experimentally indistinguishable from primary electrons. 
Such events introduce ambiguity into the $e$-$\gamma$ classification task. 
By restricting the $\gamma$ sample to pair-production events, the classifier is trained on topologies that are genuinely distinct from electrons, improving the robustness of the learning process.
The procedure WCSim uses to force pair production is to first simulate the gamma ray until it converts, and then to place both conversion products at the original primary vertex, retaining only their species and momenta. The conversion distance is not propagated. Our gamma events are therefore $e^{+}e^{-}$ pairs originating at the vertex with realistic pair-production kinematics, and the conversion gap is absent by construction. This is not a limitation for the present single-particle study, since for an isolated gamma the conversion point is unobservable in the absence of any reference marking the origin. In a full neutrino interaction, however, the displacement of the shower relative to the interaction vertex carries additional discriminating power, and exploiting it is left to future work.

Neutral pions are generated as true $\pi^{0}$ particles, which promptly decay into two photons, producing two-ring morphologies of varying opening angles relevant for classification. Unlike the gamma sample, these daughter photons are not forced to pair-produce. However, the overwhelming majority convert regardless, since pair production is the dominant photon interaction in water above $\sim\unit[10]{MeV}$ and each daughter photon carries $\sim\unit[68]{MeV}$ for a $\pi^{0}$ at rest (and more for boosted decays).

Charged pions are not included in this study. Although they can mimic muons, charged pions instead interact hadronically in the water \citep{PDG2024}, so a substantial fraction interact before stopping, producing kinked tracks and additional rings. Traditional reconstruction methods often handle this through multi-ring fits. We therefore opt not to include charged pions in this work while full multi-ring reconstruction is being investigated as an extension of this work.

\subsection{Network Architectures and Training Procedure} \label{sec:architecture}

The training, validation, and testing of all deep-learning models are carried out with the \textsc{WatChMaL} framework~\cite{watchmal2025}, which consolidates data loading, augmentation, optimization, and performance logging in a single pipeline. 
Seven distinct networks are developed. 
Three separate regression models are trained on the single-electron dataset, predicting the interaction vertex, the outgoing-track direction, and the total energy, respectively. 
An analogous trio of regression models are trained on the single-muon dataset. 
A seventh network, configured for four-class classification, is trained jointly on electrons, muons, pair-producing gamma rays, and neutral pions to provide particle-identification probabilities.

Vertex reconstruction models each produce three outputs representing the $(x,y,z)$ components of the interaction vertex. We employ the Huber loss function~\cite{huber1964}, applied element-wise to each Cartesian component, with $\delta = 20\,\mathrm{cm}$. This value was chosen to approximately match the $1\sigma$ width of the per-component vertex residual observed in preliminary training runs, placing well-reconstructed events in the quadratic regime of the loss while events in the tails are down-weighted by its linear regime.

Direction models produce three outputs corresponding to the Cartesian components $(x,y,z)$ of the particle direction. Earlier attempts to regress the polar and azimuthal angles directly yielded worse angular resolutions, so we instead regress the Cartesian components of the unit direction vector. We again employ the Huber loss, applied element-wise to the raw network outputs, with $\delta = 1$. These outputs are not constrained to unit length, so large residuals can arise early in training, and the linear regime of the Huber loss limits the influence of these outliers. The outputs are normalized to unit length only when forming the final predicted direction.

Energy reconstruction networks are trained to reconstruct the total energy $E_{\rm{total}}=T + m$ in $\rm{MeV}$ where $T$ is the kinetic energy and $m$ is the rest mass. This differs from the \textit{visible energy} discussed in Section~\ref{sec:intro}.
Momentum is then obtained via $p=\sqrt{E_{\rm{total}}^2 - m^2}$. All momentum-reconstruction results in this analysis are based on this converted quantity.

We adopt a \textit{relative Huber loss}, which replaces the absolute residual with the relative residual. Specifically, given target $E$, prediction $\hat{E}$, and threshold $\delta > 0$, the relative Huber loss is 
\begin{equation}
L(\hat{E}, E) =
\begin{cases}
\frac{1}{2} \left(\frac{\hat{E} - E}{E + \varepsilon}\right)^{2}, & \text{if } \left|\frac{\hat{E} - E}{E + \varepsilon}\right| \leq \delta \\[1.2em]
\delta \left|\frac{\hat{E} - E}{E + \varepsilon}\right| - \tfrac{1}{2}\delta^{2}, & \text{otherwise}
\end{cases} ,
\end{equation}
where $\varepsilon$ is a small constant added for numerical stability ($10^{-6}$ in our models), and $\delta$ is set to $0.1$ in this case. This formulation is adopted because using an absolute energy residual would overweight errors at high energies: for example, a $2\,$MeV discrepancy is weighted equally whether the true total energy is $4\,$MeV or $1000\,$MeV, yet the former represents a $50\%$ error while the latter represents only a $0.2\%$ error. By expressing the residual relative to the true energy, the loss appropriately reflects the fractional error across the full energy range.

The classifier uses the standard cross-entropy loss with identical optimizer settings and returns a four-component softmax distribution over the $e$, $\mu$, $\gamma$, and $\pi^{0}$ hypotheses.

Each dataset is partitioned into training, validation, and test subsets in a $(96{:}2{:}2)$ ratio, with class balance enforced for the classifier.
The $96{:}2{:}2$ split ensures that the ($\sim 2\times10^{5}$) events allocated to the validation and test sets (before selection cuts) provide sufficient statistics in each particle species and momentum bin to yield stable estimates of resolution and bias, while the remaining \(\approx 9.6\) million events per class are used for training.

All networks use the same fundamental architectural design derived from the 152-layer residual network~\cite{He2016} (ResNet-152). 
The implementation follows the canonical design of bottleneck residual blocks with shortcut connections that promote stable gradient flow. 
Input images pass through an initial $4\times4$ convolution of stride two and a $3\times3$ max-pool with stride of two, followed by four stages of residual blocks with channel depths $\{64,\,128,\,256,\,512\}$. 
An adaptive-average-pooling layer reduces the spatial dimensions, after which a fully connected layer projects to the required output dimension. Kaiming-normal weight initialization~\cite{He2015} and zero-initialized residual-branch batch norms~\cite{Ioffe2015} are applied.

The absolute time recorded in the first input channel is referenced to the trigger time $T_{0}$ and is a nuisance quantity. The kinematics of an event are determined by the relative arrival times of Cherenkov photons across the PMT array, not by where those hits fall within the readout window. We therefore make the networks invariant to a constant time offset by constraining the first convolutional layer as described in~\ref{app:hit-time-invariance}. The reconstruction then depends on differences in hit times, but is exactly independent of the absolute time origin.

ResNet-50 networks were also evaluated but yielded inferior reconstruction performance relative to ResNet-152, despite offering faster training and inference. Since our primary objective was to demonstrate the best achievable reconstruction accuracy for the Hyper-Kamiokande far detector, we report only the ResNet-152 results here. Intermediate-depth networks (ResNet-101) were not investigated, though they may offer a favorable trade-off between accuracy and computational cost; a systematic study of network depth is left to future work.

All six regression models are optimized for twenty epochs on four NVIDIA A100 GPUs using synchronous data-parallel training. The classifier is optimized for 12 epochs using the same set-up. A mini-batch of 256 events is processed per optimization step, with the batch distributed evenly across the four devices. Every 2000 iterations, training is paused and a validation phase is executed, during which 690 mini-batches of 256 events are evaluated, corresponding to $1.77\times10^{5}$ validation events per cycle. These large validation sizes were chosen to minimize statistical fluctuations in the validation loss, ensuring that the best-performing checkpoint reflects genuine improvements in network performance rather than a favorable draw of easily reconstructed events. No explicit early-stopping criterion is imposed; instead, the model parameters corresponding to the best validation performance are retained. These models are then used for all test-set evaluations in Section~\ref{sec:results}, ensuring that only the optimal checkpoint is used for evaluation even if overfitting occurs in later epochs.

All networks use a variant of the \textsc{AdamW} optimizer~\cite{Loshchilov2017} in which weight decay is applied only to parameters of rank two or higher, that is, to convolutional kernels and fully connected weight matrices. The initial learning rate is $10^{-2}$ and the weight decay $0.005$. The learning rate follows a single cosine decay spanning the entire training run, from its initial value to $\eta_{\min}=10^{-6}$. These hyperparameters were selected through iterative manual tuning rather than a formal grid search, retaining the configuration that yielded the best performance across a series of exploratory training runs.

All metrics are computed on the held-out test samples, which amount to roughly $2\times10^{5}$ single-electron events and an equal number of single-muon events for the six regression models, and to about $8.0\times10^{5}$ events for the four-class classifier (approximately $2\times10^{5}$ per particle species). 
Before ResNet-152 networks are applied, each event is required to satisfy a set of fiducial and quality criteria designed to guarantee sufficient information content and reliable containment.

An event is considered admissible for evaluation if it satisfies two conditions. First, it must register at least 200 photo-electron equivalent hits within the $1.35\,\mu\mathrm{s}$ readout window, ensuring a sufficiently strong signal above the expected $\sim 110$ dark-noise hits. Second, the event must be fully contained: any charged particle that leaves the cylindrical water volume with kinetic energy above its Cherenkov threshold causes the event to be rejected. Because the simulations do not include the outer-detector veto PMTs, this containment criterion is applied geometrically by requiring that all particle stopping points lie inside the inner-detector boundary before falling below Cherenkov threshold.

For the regression networks, only the containment requirement is applied during training; no hit number threshold is imposed at that stage. During testing, however, both the containment cut and the 200-hit cut are applied. After these selections, approximately $92.5\%$ of the electron sample and $87.9\%$ of the muon sample remain, corresponding to about $1.85\times10^{5}$ and $1.75\times10^{5}$ test events, respectively.

For the classification network, neither the containment cut nor the hit cut is applied during training so that class balance across the four particle species is preserved. Both cuts are imposed only at testing. After these evaluation selections, $92.8\%$, $92.5\%$, $87.9\%$, and $91.8\%$ of the $\gamma$, $e$, $\mu$, and $\pi^{0}$ events were retained, respectively.

\section{Reconstruction Results} \label{sec:results}

\subsection{Kinematic Reconstruction} \label{sec:regression_results}
To quantify the quality of each regressed quantity, we employ metrics that follow the conventions of earlier Super-Kamiokande and Hyper-Kamiokande analyses~\cite{SK2019IV, HK2015, Gao2024}.

Momentum accuracy is characterized by the fractional bias, defined as the mean of 
$\Delta p/p_{\mathrm{true}}$ with $\Delta p = p_{\mathrm{pred}} - p_{\mathrm{true}}$, reported together with the \emph{resolution} $\sigma_{68}\left(\frac{\Delta p}{p_{\mathrm{true}}}\right)$, defined as the 68th percentile of the absolute value of the bias of each event, i.e. the value such that $68\%$ of test events have a smaller absolute value of $\Delta p/p_{\mathrm{true}}$.
Although the network directly reconstructs the particle’s total \textit{energy}, we report momentum-based performance to maintain consistency with previous studies; momentum is obtained via the standard conversion $p=\sqrt{E^{2}-m^{2}}$ prior to analysis.

Vertex reconstruction performance is quantified using the Euclidean distance $\Delta r = \bigl\lVert\mathbf{r}_{\mathrm{pred}} - \mathbf{r}_{\mathrm{true}}\bigr\rVert_{2}$.
We define the distance resolution $\sigma_{68}\left(\Delta r\right)$ such that $68\%$ of the events have smaller $\Delta r$ values.

For direction reconstruction, we use the opening angle $\Delta\theta = \cos^{-1}\bigl(\hat{\mathbf{u}}_{\mathrm{pred}}\!\cdot\!\hat{\mathbf{u}}_{\mathrm{true}}\bigr)$ between the reconstructed and true unit vectors.
We define the angular resolution $\sigma_{68}\left(\Delta \theta\right)$ such that 68\,\% of events have smaller $\Delta\theta$ values.

\begin{figure}[ht]
 \centering
 \includegraphics[width=\linewidth]{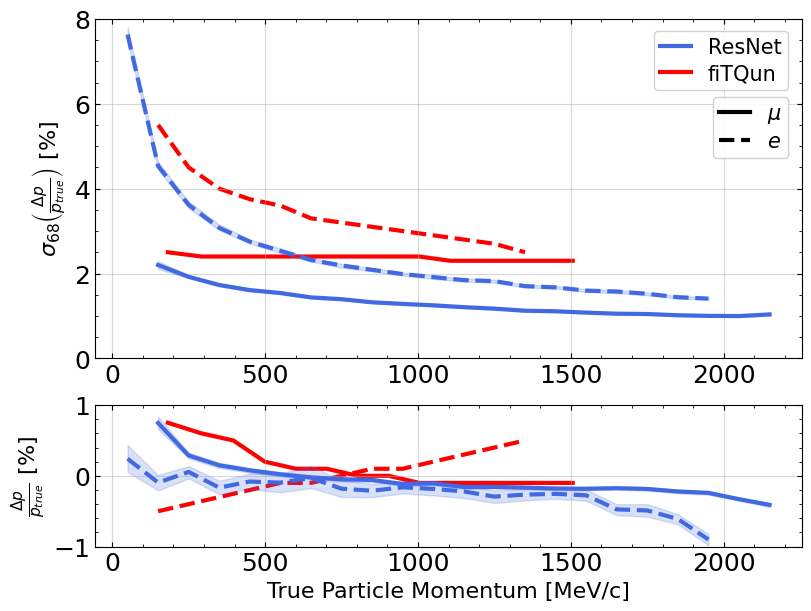}
 \caption{
 Momentum reconstruction results for muon (solid) and electron (dashed) test events using the ResNet-152 network (blue), compared with the Super-Kamiokande-IV~\cite{SK2019IV} \textsc{fiTQun} results (red). The upper panel shows the fractional momentum resolution, expressed as a percentage and defined as the 68th percentile within each 100 MeV/c bin; the lower panel shows the fractional bias, also expressed as a percentage. Shaded bands indicate $95\%$ bootstrap confidence intervals on the ResNet results, though these are narrower than the line width throughout, except for the electron bias.}
 \label{fig:mom-reg}
\end{figure}

\begin{figure}[ht]
 \centering
 \includegraphics[width=\linewidth]{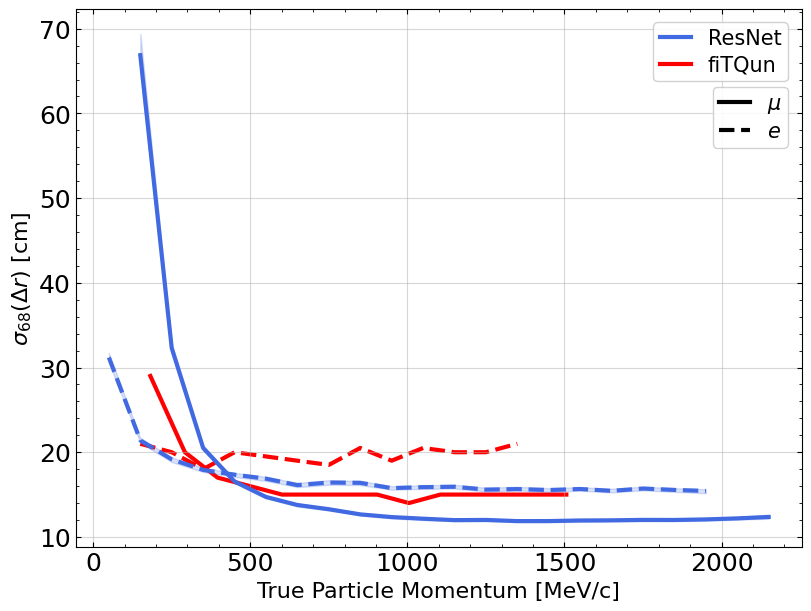}
 \caption{Vertex-reconstruction resolution for the ResNet-152 models (blue) and \textsc{fiTQun} results from Super-Kamiokande-IV~\cite{SK2019IV} (red). Reconstruction performance for muons (solid) and electrons (dashed) are shown. The resolution is defined as the three-dimensional Euclidean distance below which $68\%$ of events fall per 100 MeV/c true momentum bin. Shaded bands indicate $95\%$ bootstrap confidence intervals on the ResNet results, though these are largely smaller than the line width and are therefore barely visible.}
 \label{fig:pos-reg}
\end{figure}

\begin{figure} [ht]
 \centering
 \includegraphics[width=\linewidth]{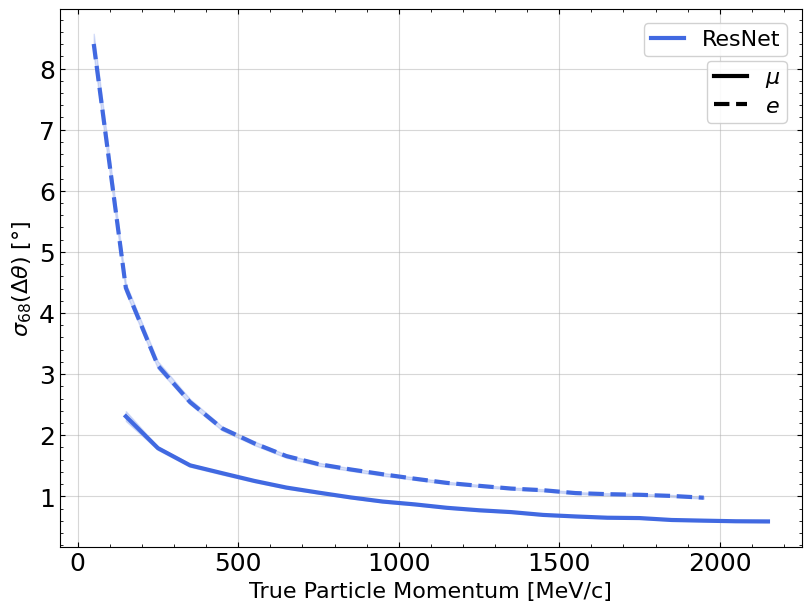}
 \caption{Direction reconstruction performance of the ResNet-152 models for muon (solid) and electron (dashed) events. The angular resolution is quoted at the 68th percentile per 100 MeV/c true momentum bin, where the opening angle $\Delta \theta$ is defined as the angle between the true and reconstructed particle directions. Shaded bands indicate $95\%$ bootstrap confidence intervals on the ResNet results; these are narrower than the line width throughout.}
 \label{fig:dir-reg}
\end{figure}

All reported resolutions, biases, and classification metrics are accompanied by $95\%$ confidence intervals estimated via bootstrap resampling. For each quantity, 1000 resamples of the test set are drawn with replacement, and the metric is recomputed on each resample. The $95\%$ confidence interval is then taken as the 2.5th and 97.5th percentiles of the resulting distribution. Resampling is performed independently within each momentum bin for Figures~\ref{fig:mom-reg},~\ref{fig:pos-reg},~\ref{fig:dir-reg}, and on the full test sample for the summary statistics in Tables~\ref{tab:reg-summary} and~\ref{tab:classification-summary}. Given the large test sample sizes ($\sim$$1.75\times10^{5}$--$1.85\times10^{5}$ events), the resulting intervals are narrow throughout, confirming that the precision of the reported metrics is statistically justified.  An exception is the electron signal efficiency shown in Figures~\ref{fig:PID-signal-efficiency},~\ref{fig:PID-signal-efficiency-zoomed}, and~\ref{fig:PID-signal-efficiency-small-dwall/towall}, for which error bars represent the binomial standard error on the efficiency in each bin, as is conventional for proportion estimates.

Throughout this work, all resolution and efficiency profiles are shown as functions of true particle momentum, following common practice in Super-Kamiokande and Hyper-Kamiokande reconstruction studies and allowing direct comparison with existing results. For reference, we also include the reported \textsc{fiTQun} performance from the Super-Kamiokande-IV atmospheric oscillation analysis~\cite{SK2019IV}, shown in red in Figures~\ref{fig:mom-reg} and~\ref{fig:pos-reg}. 
It is important to stress, however, that differences in detector geometry, event selection, and \textsc{fiTQun} configuration render this comparison only approximate; a definitive evaluation would require running \textsc{fiTQun} on the identical simulated events used in this work, which falls within the remit of the Hyper-Kamiokande Collaboration.

The Super-Kamiokande-IV study also reported some reconstruction results as a function of \emph{visible energy}, i.e. the energy of an electromagnetic shower that would produce the observed amount of Cherenkov light.
This quantity is detector-specific and is not the same as the true total energy, kinetic energy, or momentum of the primary particle reconstructed in our study. For Figures~\ref{fig:mom-reg} and~\ref{fig:pos-reg}, we therefore convert the published visible-energy values to corresponding true momenta to enable a more direct, though still approximate, comparison with our momentum-dependent resolutions.

Average resolutions and biases for each kinematic regression task are summarized in Table~\ref{tab:reg-summary}, along with reported values from the Super-Kamiokande IV study~\cite{SK2019IV} and the Hyper-Kamiokande physics-potential study~\cite{HK2015}, where applicable for the muon and electron regression tasks.

\begin{table*}[t]
    \centering
\begin{tabular}{ll|ccc|ccc}
    \toprule
    & & \multicolumn{3}{c}{$\mu$} & \multicolumn{3}{c}{$e$} \\
    \cmidrule(lr){3-5} \cmidrule(lr){6-8}
    & & ResNet & SK-IV & HK 2015 & ResNet & SK-IV & HK 2015 \\
    \cmidrule(lr){1-2} \cmidrule(lr){3-8}
    \multirow{2}{*}{Momentum}
    & Mean Bias
        & $-0.087 \pm 0.007\%$
        & $-0.18\%$ & \textemdash{}
        & $-0.233 \pm 0.023\%$
        & $0.43\%$  & \textemdash{} \\[4pt]
    & Mean Resolution
        & $1.266 \pm 0.006\%$
        & $2.26\%$  & $2.6\%$
        & $2.209 \pm 0.012\%$
        & $2.90\%$  & $4.0\%$ \\
    \midrule
    Position
    & Mean Resolution
        & $13.63 \pm 0.04\,\mathrm{cm}$
        & $15.8\,\mathrm{cm}$ & $30\,\mathrm{cm}$
        & $17.03 \pm 0.05\,\mathrm{cm}$
        & $20.6\,\mathrm{cm}$ & $27\,\mathrm{cm}$ \\
    \midrule
    Direction
    & Mean Resolution
        & $0.908 \pm 0.004^{\circ}$
        & $1.00^{\circ}$ & \textemdash{}
        & $1.595 \pm 0.008^{\circ}$
        & $1.48^{\circ}$ & \textemdash{} \\
    \bottomrule
\end{tabular}
    \caption{Summary statistics for the regression tasks. Position and direction resolutions, together with the average momentum resolution and bias, are compared to \textsc{fiTQun} performance evaluated on CCQE single-ring events with visible energy of $\unit[1]{GeV}$ from the Super-Kamiokande-IV oscillation analysis (SK-IV)~\cite{SK2019IV}, as well as to the 500\,MeV/$c$ benchmarks reported in the 2015 Hyper-Kamiokande physics potential study (HK)~\cite{HK2015}. Uncertainties on the ResNet values represent $95\%$ bootstrap confidence intervals and are estimated using the bootstrap resampling across the entire test set.}
    \label{tab:reg-summary}
\end{table*}

Figure~\ref{fig:mom-reg} shows the fractional momentum resolution (upper panel) and mean fractional bias (lower panel) obtained with the ResNet-152 regressors (blue curves) for both muon (solid) and electron (dashed) test events. Each point on the curves corresponds to a $\unit[100]{MeV/c}$ bin in true momentum; however, the markers themselves are omitted for visual clarity.

For muon momentum reconstruction, the neural network achieves a smaller bias and better resolution than the Super-Kamiokande-IV \textsc{fiTQun} results across the full momentum range shown, though noting the limitations of this indirect comparison. The momentum bias remains below $0.5\%$ for all particle momenta above $\unit[250]{MeV/c}$, while the resolution improves from about $\sim2.1\%$ at $\unit[150]{MeV/c}$ to $\sim1\%$ at $\unit[2.2]{GeV/c}$.

Electron momentum reconstruction shows a similar trend. The ResNet-152 model maintains a uniformly low bias and achieves a resolution that decreases from about $5\%$ at $\unit[150]{MeV/c}$ to $1.5\%$ at $\unit[2]{GeV/c}$. 
The bias hovers between $0.5\%$ and $-0.5\%$ across most of the momentum range, except at higher momenta where it climbs to $-1\%$.
Within the same caveats, this approach achieves electron momentum resolutions that are better than the Super-Kamiokande-IV \textsc{fiTQun} results over the kinematic range relevant to single-ring electron analyses.

Networks trained to reconstruct the interaction vertex achieve average resolutions of approximately $13.6\,\mathrm{cm}$ for muons and $17.0\,\mathrm{cm}$ for electrons (Table~\ref{tab:reg-summary}). Figure~\ref{fig:pos-reg} shows the resolution as a function of true momentum, alongside \textsc{fiTQun} results from the Super-Kamiokande-IV oscillation analysis in red~\cite{SK2019IV}. Similar to momentum reconstruction, performance for both electron and muon regressors degrades noticeably at lower momenta, reflecting the reduced hit multiplicity in these events. In spite of this, our models achieve better position resolutions across most of the momentum range for both particle types.

The direction-regression networks yield mean angular resolutions of $0.91^{\circ}$ and $1.60^{\circ}$ for muon and electron events respectively.
However, degradation near threshold momenta arises due to reduced photon statistics and increased multiple Coulomb scattering, which smear the Cherenkov ring, diminishing the directional information carried by the timing and charge patterns.

Taken together, these results show that the ResNet-based reconstruction models achieve kinematic resolutions that are better than the maximum-likelihood fitter \textsc{fiTQun} across most of the momentum range, while delivering substantially faster inference (see Section~\ref{sec:discussion} for details on computational speed). 

\subsection{Particle Classification} \label{sec:classification_results}

At $\mathcal{O}(\mathrm{GeV})$, a photon in water interacts almost exclusively through pair production, yielding an $e^{+}e^{-}$ pair with an opening angle typically below $1^{\circ}$~\cite{PDG2024}. 
Consequently, the Cherenkov signature of a photon-initiated shower is, to first order, indistinguishable from that of a primary electron. 
Neutral pions, in contrast, promptly decay via $\pi^{0}\!\rightarrow\!\gamma\gamma$.
The two daughter photons are usually separated by $\mathcal{O}(10^{\circ})$, producing a broader electromagnetic pattern~\cite{PDG2024}.
These similarities complicate the definition of \emph{signal} and \emph{background} when the network simultaneously assigns probabilities to four classes $\{e,\,\mu,\,\gamma,\,\pi^{0}\}$. To create receiver operating characteristic (ROC) curves from the softmax output $\{P_e,P_\mu,P_\gamma,P_{\pi^{0}}\}$ we therefore define tailored binary discriminants, each normalized to the phase space relevant for the physics question at hand.

\begin{figure}[ht]
    \centering
    \includegraphics[width=1\linewidth]{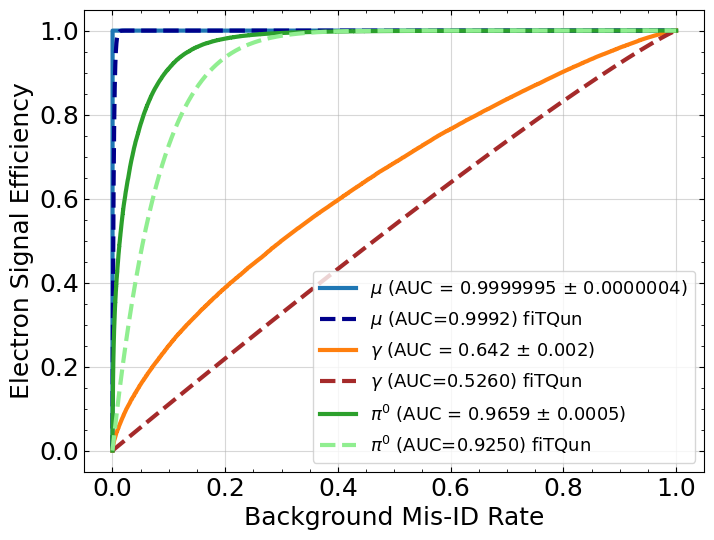}
    \caption{ROC curves for electron signal versus background from other particle species. The curves correspond to $e$-$\mu$ (blue), $e$-$\gamma$ (orange), and $e$-$\pi^{0}$ (green) classification. For comparison, the performance of \textsc{fiTQun} reported in Gao et al.~\cite{Gao2024} is shown with dashed lines, while solid lines indicate the performance of our model. 
    The AUC for each curve is listed in the legend along with its $95\%$ confidence interval. Discriminators used are defined in Eqs.~\ref{eq:Demu}-\ref{eq:Degamma}.
    Note that the \textsc{fiTQun} comparisons should be treated as indicative rather than definitive, as the configuration and cuts used in Gao et al. may not reflect the optimal \textsc{fiTQun} performance for each task.}
    \label{fig:roc-e-vs-rest}
\end{figure}

For the $e$-vs-$\mu$ ROC (blue in Fig.~\ref{fig:roc-e-vs-rest}) we treat the entire electromagnetic phase space ($e,\,\gamma,\,\pi^{0}$) as \emph{signal}. The corresponding discriminant is then 
\begin{equation}
   D_{e/\mu} \;=\; P_e + P_\gamma + P_{\pi^{0}} \;=\; 1 - P_\mu.
   \label{eq:Demu}
\end{equation}
The complementary discriminator for $\mu$ signal against an electron background is simply $D_{\mu/e}=P_\mu$.

For the $e$-vs-$\pi^{0}$, since the neutral pions are more readily separable from single electromagnetic showers than muons are, we refine the numerator to include only the photon-like subclasses, while normalizing to the $(e,\gamma,\pi^{0})$ subspace: 
\begin{equation}
   D_{e/\pi^{0}}
   \;=\;
   \frac{P_e + P_\gamma}{P_e + P_\gamma + P_{\pi^{0}}}
   \;=\;
   1 - \frac{P_{\pi^{0}}}{P_e + P_\gamma + P_{\pi^{0}}}.
   \label{eq:Depi0}
\end{equation}
Finally, discriminating between electrons and photons requires maximal sensitivity to their subtle topological differences. We therefore normalize to the $(e,\gamma)$ subspace only:
\begin{equation}
   D_{e/\gamma}
   \;=\;
   \frac{P_e}{P_e + P_\gamma}
   \;=\;
   1 - \frac{P_\gamma}{P_e + P_\gamma}.
   \label{eq:Degamma}
\end{equation}

The selection of these discriminants ensures that each ROC curve probes the intrinsic separability of the targeted particle hypotheses without being diluted by classes that are either obviously distinct (muons in the $e$-$\pi^{0}$ study) or almost completely degenerate (photons in the $e$-$\mu$ study). Figure~\ref{fig:roc-e-vs-rest} summarizes the ROC curves for electron(-like) efficiency vs background of each particle class. 
All ROC curves in this work are computed from the corresponding $D$ values using threshold scans over the interval $[0,1]$.

Across the entire kinematic range the network attains electron(-like) vs background AUC values of $0.9999995\pm0.0000005$ ($e$-$\mu$), $0.642 \pm 0.002$ ($e$-$\gamma$), and $0.9659 \pm 0.0005$ ($e$-$\pi^{0}$). For comparison, the \textsc{fiTQun} AUC results reported in Gao et al.~\cite{Gao2024} are also shown as dashed lines in Fig.~\ref{fig:roc-e-vs-rest}, and our ResNet models outperform \textsc{fiTQun} across all three tasks. We note, however, that two caveats apply to this comparison. First, a dedicated $e$-$\gamma$ separator has not previously been achieved for likelihood-based reconstruction, so the \textsc{fiTQun} $e$-$\gamma$ result in Gao et al. should be interpreted with caution and the comparison treated as indicative rather than definitive. Second, the configuration and cuts used in Gao et al. may not reflect the optimal \textsc{fiTQun} performance for the $e$-$\pi^{0}$ task, and the comparison should likewise be treated as indicative rather than definitive.

\begin{figure}[ht]
    \centering
    \includegraphics[width=1\linewidth]{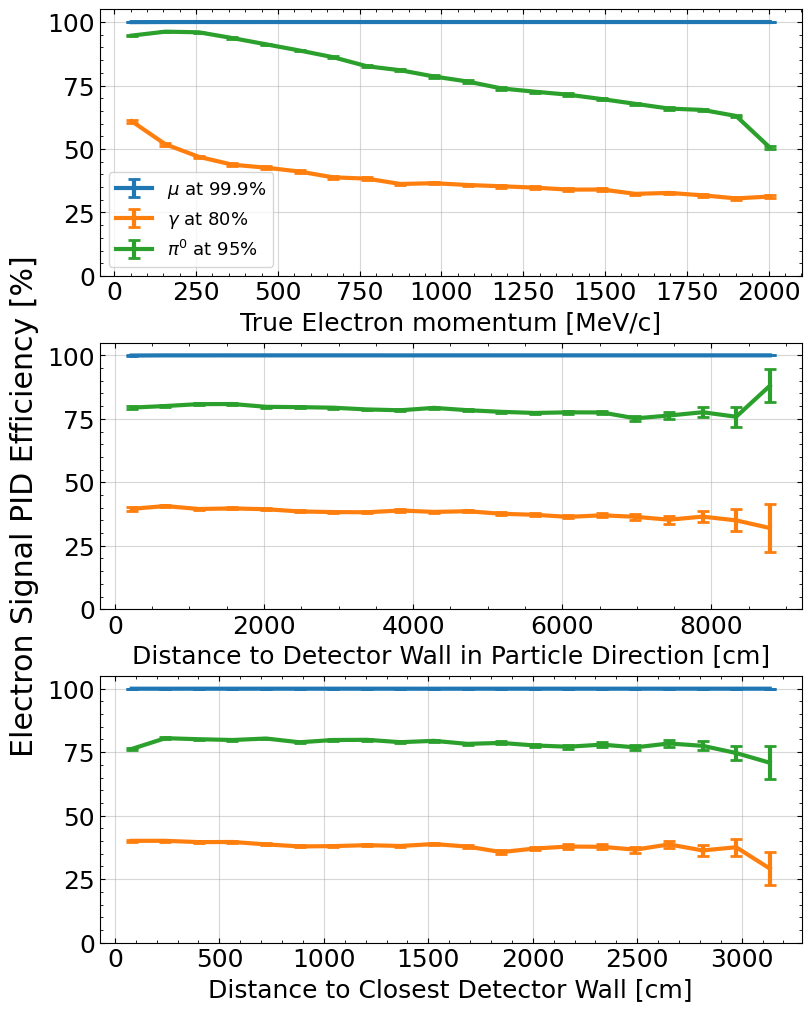}
    \caption{Electron selection efficiency at fixed background-rejection operating points of $99.9\%$ for $\mu$, $80\%$ for $\gamma$, and $95\%$ for $\pi^{0}$ candidates. The blue curve corresponds to the $(e+\gamma+\pi^{0})$ signal versus $\mu$ background discriminator, the green curve to the $(e+\gamma)$ signal versus $\pi^{0}$ background, and the orange curve to the $e$ signal versus $\gamma$ background, as defined in Eqs.~\ref{eq:Demu}-\ref{eq:Degamma}. The top panel shows the efficiency as a function of true electron momentum; the middle panel as a function of the distance to the detector wall along the particle direction; and the bottom panel as a function of the distance from the interaction vertex to the nearest detector wall. Error bars show the binomial standard error on the efficiency in each bin. Zoomed views of the near-wall performance are shown in Fig.~\ref{fig:PID-signal-efficiency-small-dwall/towall} in~\ref{sec:extended-diagnostics}.}
    \label{fig:PID-signal-efficiency}
\end{figure}

We parse these results as functions of true momentum and of the distance to the inner-detector wall in the particle's direction of travel in Fig.~\ref{fig:PID-signal-efficiency}.

This study follows the convention adopted for the Intermediate Water Cherenkov Detector, where classifier performance is expressed in terms of the electron particle-ID efficiency required to achieve background rejections of $99.9\%$ for $\mu$, $80\%$ for $\gamma$ candidates~\cite{Prouse2023}, and $95\%$ for $\pi^{0}$. For each task, a single discriminator threshold is determined globally from the full test sample by finding the value at which the desired fraction of background events are rejected. This threshold is then applied uniformly across all events, and the electron signal efficiency in each bin is computed as the fraction of true electron events whose discriminator value exceeds the threshold.

Accordingly, Figure~\ref{fig:PID-signal-efficiency} shows the fraction of true electrons retained under each of these background-rejection operating points. The top panel presents this efficiency as a function of true electron momentum, the middle panel as a function of the distance from the interaction vertex to the detector wall along the particle direction, and the bottom panel as a function of the distance from the interaction vertex to the nearest detector wall. The latter two metrics are included as traditional reconstruction is known to degrade for near-wall events, where the Cherenkov cone intersects the detector wall at very short distances, causing most of the light to be collected by only a few PMTs and making the ring pattern difficult to resolve.

Table~\ref{tab:classification-summary} presents performance metrics computed using the signal-background definitions defined for Eqs.~\ref{eq:Demu}-\ref{eq:Degamma}.   
Purity, efficiency, accuracy, and F1-scores are evaluated at the same fixed background-rejection operating points used in the efficiency profiles, namely $99.9\%$ for $\mu$, $95\%$ for $\pi^{0}$, and $80\%$ for $\gamma$ candidates.

\begin{table}[t]
\centering
\small
\setlength{\tabcolsep}{4pt}
\renewcommand{\arraystretch}{1.05}     

\begin{tabular}{c||c||ccc|c}
\toprule
Task & Class & Purity & Efficiency & F1-score & Accuracy \\
\midrule
\multirow{2}{*}{$e$ vs.\ $\mu$}
  & $e+\gamma+\pi^{0}$ & 0.9997 & 1.0000 & 0.9998 & \multirow{2}{*}{0.9997} \\
\cmidrule(lr){2-5}
  & $\mu$              & 0.9999 & 0.9990 & 0.9995 & \\
\midrule
\multirow{2}{*}{$e$ vs.\ $\pi^{0}$}
  & $e+\gamma$         & 0.969  & 0.777  & 0.863  & \multirow{2}{*}{0.835} \\
\cmidrule(lr){2-5}
  & $\pi^{0}$          & 0.679  & 0.950  & 0.792  & \\
\midrule
\multirow{2}{*}{$e$ vs.\ $\gamma$}
  & $e$                & 0.659  & 0.388  & 0.488  & \multirow{2}{*}{0.594} \\
\cmidrule(lr){2-5}
  & $\gamma$           & 0.567  & 0.800  & 0.664  & \\
\bottomrule
\end{tabular}

\caption{ResNet classification metrics evaluated using the discriminants defined in Eqs.~\ref{eq:Demu}-\ref{eq:Degamma}. Purity, efficiency, F1-score, and accuracy are reported at fixed background-rejection operating points: $99.9\%$ for $\mu$, $95\%$ for $\pi^{0}$, and $80\%$ for $\gamma$, matching those used in the efficiency-profile comparison (Fig.~\ref{fig:PID-signal-efficiency}). "Task" denotes the specific binary classification problem between an electron(-like) signal and a chosen particle background, while "Class" refers to the two labels used within that task, namely the signal class and the background class. For example, in the $e$ vs. $\mu$ task, the signal class corresponds to the combined $e+\gamma+\pi^{0}$ sample, and the background class corresponds to $\mu$.}
\label{tab:classification-summary}
\end{table}

\section{Discussion} \label{sec:discussion}

This work demonstrates that ResNet-based networks achieve classification and regression resolutions that are better than those of the established maximum-likelihood algorithm \textsc{fiTQun}, while offering substantially reduced inference time.

To evaluate the computational performance of the proposed models, we compare inference times against the traditional likelihood-based reconstruction algorithm \textsc{fiTQun}. Using a single NVIDIA A100 ($\unit[80]{GB}$) GPU, our ResNet models process $1.75\times10^{5}$ simulated muon events in $\unit[235]{s}$, corresponding to an average inference time of $\unit[1.3]{ms}$ per event. For electrons, $1.85\times10^{5}$ events are processed in $\unit[316]{s}$, yielding an average inference time of $\unit[1.7]{ms}$ per event.

In contrast, reconstructing the same event samples with \textsc{fiTQun} requires substantially longer wall times. For muons, the average reconstruction time is $\unit[67.1]{s}$ per event ($\approx1.17\times10^{7}$ CPU seconds), while for electrons the corresponding time is $\unit[54.1]{s}$ per event ($\approx 1.00\times10^{7}$ CPU seconds). These timings include the full \textsc{fiTQun} reconstruction chain, comprising pre-fitting and clustering, single-ring likelihood fits (particle ID, vertex, energy, and direction reconstruction), and $\pi^{0}$ hypothesis fits. Overall, this corresponds to speed-ups of approximately $\sim5.2\times10^{4}$ for muons and $\sim3.2\times10^{4}$ for electrons when using ResNet-based inference relative to \textsc{fiTQun}.

Such orders-of-magnitude gains in throughput are consistent with previous machine-learning studies in water Cherenkov detectors~\cite{Aguilar2025, Prouse2023} and render deep-learning reconstruction well suited for large-scale Monte Carlo production, systematic ensemble studies, and calibration workflows required by Hyper-Kamiokande.

Across all kinematic regression tasks, reconstruction performance remains within desirable regimes but degrades as the true track momentum approaches the Cherenkov threshold. The Cherenkov light yield scales with $1 - 1/(\beta^{2}n^{2})$~\cite{Frank1937}, and therefore drops rapidly near threshold, producing only a few tens of prompt photons rather than the thousands observed at the GeV scale. In this extremely low-light-yield region, even a small number of dark-noise hits falling within the narrow prompt time window can become comparable to the signal, degrading the vertex, angular, and momentum reconstructions. This dark-noise-limited behavior is confined to within a few tens of MeV/c above threshold (roughly $E \lesssim 50$ MeV for electrons and $E \lesssim 170$ MeV for muons). At higher momenta, where events produce hundreds of photons, dark noise becomes negligible; the remaining degradation arises primarily from limited photon statistics, as too few prompt photons are available to define a sharp Cherenkov ring. In this statistics-limited regime, reconstruction uncertainties scale approximately as $1/\sqrt{N_{\mathrm{hits}}}$.

The classification network achieves excellent $\mu$-$e$ separation across nearly the entire kinematic phase space. We observe near-perfect discriminability as a function of both true momentum and distance-to-wall in the particle direction. Only a modest decline in separation power appears for events originating extremely close to the detector wall, a well-known failure mode in water-Cherenkov detectors with sparse photocoverage such as Hyper-Kamiokande. In this regime, the charged lepton is effectively emitted directly in front of a PMT, causing most Cherenkov photons to be collected by a single tube. The resulting loss of angular information limits the model’s performance by the intrinsic photocoverage. While a fiducial-volume cut applied prior to reconstruction would naturally improve overall classification performance, we show that it is not required for the network to maintain strong discriminating power, even in this extreme configuration.

High separability is also achieved for the $e$-$\pi^{0}$ classification task. In Figure~\ref{fig:PID-signal-efficiency}, the electron signal efficiency remains relatively strong despite the stringent requirement of $95\%$ $\pi^{0}$ background rejection. The efficiency decreases monotonically with increasing momentum, likely due to the increasing boost of the neutral pion at higher energies. As the $\pi^{0}$ becomes more boosted, the opening angle of its decay photons narrows, eventually producing nearly collinear $\gamma\gamma$ pairs. Such topologies are increasingly gamma-like, and therefore more electron-like, eroding the separability between electrons and $\pi^{0}$ decay products. In addition, higher-energy electromagnetic showers tend to be more spatially diffuse and complex, further complicating the discrimination task.

Figure~\ref{fig:PID-signal-efficiency} also shows that the electron signal efficiency remains largely flat even at small distance-to-wall values in the particle direction, an event topology that has historically challenged likelihood-based algorithms. There is, however, a small decline in performance at large distances to the wall along the particle direction, most notably in the $e$-$\pi^{0}$ discrimination. This could possibly be attributed to greater photon attenuation as the travel distance of the photons increases. It could also be an artifact of the large statistical variance observed at distances $\gtrsim \unit[8000]{cm}$, which arises from limited bin statistics: because vertices were uniformly distributed with isotropic directions, only a small fraction of events can achieve such large distances (e.g., those originating near one extreme of the tank and pointing toward the opposite end). Many more geometric configurations populate lower distance-to-wall bins, naturally reducing variance there.

Distinguishing electrons from gamma rays in the $e$-$\gamma$ task remains intrinsically challenging, as pair-produced gammas can closely mimic primary electrons, requiring the network to exploit subtle differences in event topology. Despite this difficulty, our ResNet models achieve non-trivial $e$-$\gamma$ separation. The discriminant distributions for the two classes are distinguishable at the population level (AUC $=0.642\pm0.002$), even though per-event classification remains weak. To our knowledge this has not previously been demonstrated in this geometry. As such, the resulting resolving power significantly exceeds that of \textsc{fiTQun} and is comparable to the best machine-learning results obtained in other detector contexts~\cite{Prouse2023}. Further study to investigate exactly which features the network learned that give it its $e$-$\gamma$ discriminating power would be informative.

As in the $e$-$\pi^{0}$ case, the electron signal efficiency declines with increasing momentum, possibly due to increasingly complex electromagnetic showers, though this effect warrants further study. The dependence on distance-to-wall is similarly consistent with the $e$-$\pi^{0}$ behavior, with increased variance at $\gtrsim\unit[8000]{cm}$ attributable to the same geometric and statistical limitations described above.

The reconstruction performance achieved by our networks directly mitigates several of the challenges outlined in the introduction. Their millisecond-scale inference time removes the dominant computational burden associated with likelihood-based reconstruction, making it feasible to process the very large, systematically varied Monte Carlo ensembles required for precision oscillation analyses in Hyper-Kamiokande. Furthermore, the networks exhibit stable performance in geometries where likelihood fits are known to struggle, such as near-wall vertices or events with limited photon statistics, demonstrating improved robustness in topologies that previously demanded aggressive fiducial-volume cuts or manual quality selections. Together, these results suggest that machine-learning reconstruction can substantially reduce both the computational cost and operational complexity of large-scale Hyper-Kamiokande analysis campaigns.

\section{Conclusion} \label{sec:conclusion}
In this study we have presented the first application of deep-learning-based event reconstruction to the Hyper-Kamiokande far-detector geometry. Using simulated single-particle samples, we trained an ensemble of ResNet-152 networks for four-class particle identification and for regression of the lepton momentum, direction, and interaction vertex for electrons and muons up to 2~GeV.

Within the limitations of an indirect comparison to existing Super-Kamiokande-based \textsc{fiTQun} results, the kinematic resolutions achieved by our networks are better than those of traditional likelihood methods across most of the relevant momentum range. At the same time, the ResNet models deliver per-event inference times of $\mathcal{O}(\mathrm{ms})$ on a single GPU, corresponding to speed-ups of $3.2\times10^{4}$-$5.2\times10^{4}$ relative to typical likelihood-based reconstruction. This directly addresses a key computational bottleneck for Hyper-Kamiokande, where precision oscillation analyses require very large, systematically varied Monte Carlo samples and ensembles of pseudo-experiments.

The combination of comparable resolution and orders-of-magnitude faster inference opens up new analysis possibilities. In particular, it becomes feasible to reconstruct large numbers of MC samples under varying flux, cross-section, and detector-response assumptions, and to propagate these systematics through to reconstructed observables without prohibitive CPU cost. This capability is essential for future studies of $\nu_{\mu}\rightarrow\nu_{e}$ appearance, $\delta_{\mathrm{CP}}$ sensitivity, and common backgrounds such as neutral-current $\pi^{0}$ production.

Taken together, these results show that deep-learning-based reconstruction can serve as a practical and powerful complement to established likelihood techniques in Hyper-Kamiokande. Traditional algorithms such as \textsc{fiTQun} will remain important for detailed calibrations and cross-checks, while machine-learning models offer a scalable tool for large-scale Monte Carlo production and systematic studies. Extending these methods to more complex event topologies, such as multi-ring events 
arising from neutral-current $\pi^{0}$ production and multi-particle final states, and improving the robustness of kinematic reconstruction against detector and interaction systematic uncertainties will be natural next steps toward a fully operational machine-learning reconstruction pipeline for Hyper-Kamiokande.

\section{Acknowledgments}

A.A., E.T., and P.U. are supported by the Australian Research Council DP250100373.
E.T. is additionally supported by CE230100016.

P.D. and N.P. were supported by the Government of Canada’s New Frontiers in Research Fund [NFRFE-2019-00278] for the initial development of WatChMaL.

For N.P. support was provided by Schmidt Sciences, LLC and the UKRI Horizon Europe MSCA Guarantee Postdoctoral Fellowship EP/X027368/1.

This work has been performed in the context of the Hyper-Kamiokande Collaboration and is supported by its membership.

This work was performed on the OzSTAR national facility at Swinburne University of Technology. The OzSTAR program receives funding in part from the Astronomy National Collaborative Research Infrastructure Strategy (NCRIS) allocation provided by the Australian Government.

\bibliographystyle{elsarticle-num}
\bibliography{inline-edits-main}

\appendix

\section{Physics Motivation} \label{sec:physics-background}

Understanding the matter-antimatter asymmetry of the Universe~\cite{Cannetti2012} remains one of the foremost challenges in modern physics and cosmology. Any successful baryogenesis mechanism must violate charge-parity (CP) symmetry, as one of the three famous "Sakharov conditions"~\cite{Sakharov1967,Sakharov1991}.
Although the standard model contains a CP-odd phase in the quark mixing matrix, the resulting asymmetry is at least ten orders of magnitude smaller than what is required to explain the observed baryon-to-photon ratio~\cite{PDG2024}. 
This implies that additional CP-violating physics is required.

One well-motivated proposal is thermal leptogenesis~\cite{Fukugita1986}, which converts a lepton asymmetry generated by heavy Majorana neutrino decays into a baryon asymmetry through electroweak sphalerons. The viability of this mechanism depends on CP violation in the lepton sector.
The discovery of neutrino oscillations by Super-Kamiokande~\cite{SK1998} established that the flavor of a neutrino oscillates as it propagates through space. Such oscillations can occur only if neutrinos have mass, providing the first clear evidence of physics beyond the standard model. 
Neutrino oscillations are described by the Pontecorvo-Maki-Nakagawa-Sakata matrix~\cite{Maki1962,Pontecorvo1968}, which contains a Dirac phase \( \delta_\text{CP} \) analogous to the Kobayashi-Maskawa phase in the quark sector~\cite{Kobayashi1972}. 
A non-zero \( \delta_\text{CP} \) implies that neutrino oscillations are CP-violating, which could be a low-energy manifestation of the high-energy CP violation required for leptogenesis. 
The magnitude of CP violation in neutrino oscillations is governed by the Jarlskog invariant, $J_{\text{CP}} = \tfrac{1}{8}\sin 2\theta_{12}\sin 2\theta_{23}\sin 2\theta_{13}\cos \theta_{13}\sin \delta_{\text{CP}}$, which quantifies the size of CP-violating effects in the three-flavor neutrino mixing framework.

Long-baseline accelerator experiments such as T2K~\cite{T2K2011} and NO$\nu$A~\cite{Nova2018,Nova2019} are primarily sensitive to the CP-violating phase~$\delta_{\text{CP}}$ and the mixing angle~$\theta_{23}$. 
These experiments compare the spectra of muon-neutrino disappearance and electron-neutrino appearance.
Recent combined analysis weakly favors values near \(-\pi/2\), though $\delta_\text{CP}=0$ is not yet ruled out with five-sigma significance~\cite{Esteban2024,T2K2023,Abubakar2025}. 
Precision measurement of leptonic CP violation likely requires a larger fiducial mass and higher beam luminosity than currently available. 
This motivates the upcoming Hyper-Kamiokande experiment, which will serve as the successor to the highly successful Super-Kamiokande experiment.

\section{Traditional Event Reconstruction in Water Cherenkov Detectors} \label{sec:fitqun}

The current state-of-the-art reconstruction algorithm deployed in Hyper-Kamiokande is the maximum-likelihood estimator \textsc{fiTQun}, originally developed for the MiniBooNE experiment and adapted for Super-Kamiokande~\cite{Patterson2009,missert2017}. 
The method evaluates the likelihood given by

\begin{align}
 L(\mathbf{x}) = \prod_{j}^\text{unhit} P_j(\text{unhit}|\mathbf{x}) \prod_{i}^\text{hit} P_{i}(\text{hit}|\mathbf{x}) f_q(q_i|\mathbf{x}) f_t(t_i|\mathbf{x}) ,
\end{align}
where the hypothesis vector $\mathbf{x}$ encodes the interaction vertex $(x,y,z)$, interaction time, track directions and momenta, and, where relevant, additional kinematic parameters such as track segmentation and energy loss~\cite{Abe2018}. 
The indices $j$ and $i$ run over PMTs that did not and did register a hit, respectively; $P_j$ and $P_i$ denote the corresponding hit probabilities, while $f_q$ and $f_t$ are the probability density functions for the observed charge $q_i$ and time $t_i$ of each hit PMT.

The optimal event description is obtained by minimizing $-\ln L(\mathbf{x})$ and treating discrete quantities such as particle identification (PID) through explicit hypothesis testing. 
Although \textsc{fiTQun} performs reliably for simple single-ring topologies, its accuracy degrades in the presence of multiple overlapping rings or closely spaced electromagnetic showers that are expected to be commonplace in the larger Hyper-Kamiokande detector. 
Under such circumstances, the algorithm exhibits reduced resolution and increased systematic uncertainties, and its computational demands become prohibitive.

\section{Extended Reconstruction Diagnostics} \label{sec:extended-diagnostics}

\begin{figure}[!htbp]
    \centering
    \includegraphics[width=\linewidth]{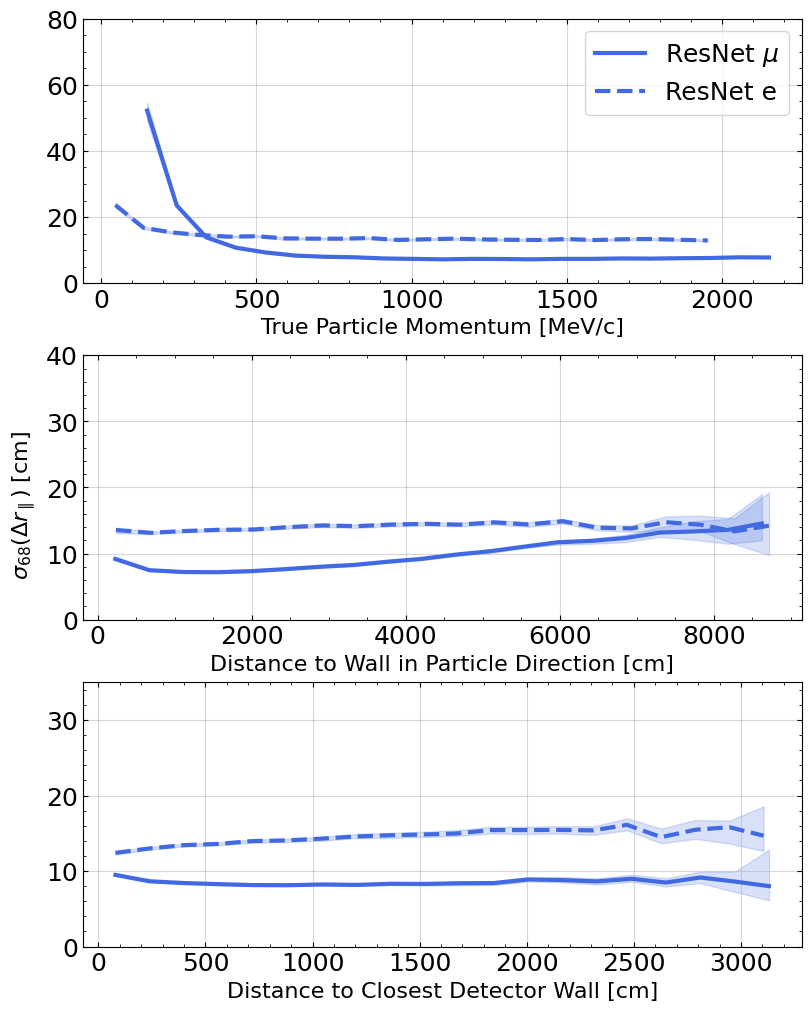}
    \caption{$\sigma_{68}(\Delta r_{\parallel})$, the 68th percentile of the longitudinal (parallel-to-track) component of the vertex displacement $\Delta r$ between the true and reconstructed interaction vertices, as a function of the true particle momentum (top), the distance to the detector wall along the particle direction (middle), and the distance to the nearest detector wall (bottom). Results for muons are indicated by solid lines, while those for electrons are indicated by dashed lines. Shaded bands indicate $95\%$ bootstrap confidence intervals, though these are narrower than the line width across most of the momentum range.}
    \label{fig:pos-long}
\end{figure}

\begin{figure}[!htbp]
    \centering
    \includegraphics[width=\linewidth]{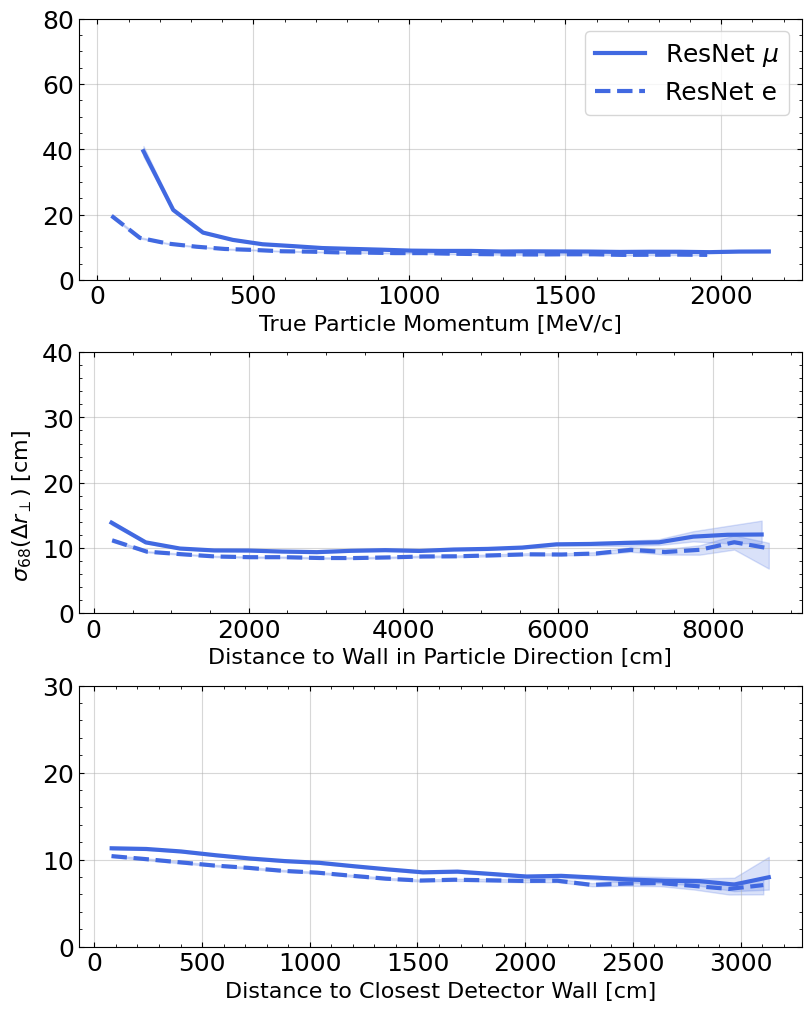}
    \caption{$\sigma_{68}(\Delta r_{\perp})$, the 68th percentile of the transverse (perpendicular-to-track) component of the vertex displacement $\Delta r$ between the true and reconstructed interaction vertices, as a function of the true particle momentum (top), the distance to the detector wall measured along the particle direction (middle), and the distance to the nearest detector wall (bottom). Results for muons are indicated by solid lines, while those for electrons are indicated by dashed lines. Shaded bands indicate $95\%$ bootstrap confidence intervals, though these are narrower than the line width across most of the momentum range.}
    \label{fig:pos-trans}
\end{figure}

Here we provide additional performance breakdowns for readers interested in more detailed diagnostics. In particular, we decompose the vertex reconstruction displacement vector $\Delta r = \bigl\lVert\mathbf{r}_{\mathrm{pred}} - \mathbf{r}_{\mathrm{true}}\bigr\rVert_{2}$ into two geometrically distinct components: the longitudinal component $\Delta r_{\parallel}$, projected onto the true particle direction, which captures errors along the track axis where the Cherenkov emission originates; and the transverse component $\Delta r_{\perp}$, projected onto the plane perpendicular to the particle direction, which captures errors in the lateral positioning of the vertex. In both cases, the resolution is quantified by $\sigma_{68}$, the 68th percentile of the absolute component error within each bin, which serves as a robust analogue of the standard deviation for potentially non-Gaussian distributions. Figure~\ref{fig:pos-long} shows $\sigma_{68}(\Delta r_{\parallel})$ as a function of the true particle momentum, the distance to the detector wall along the particle direction, and the distance to the nearest detector wall. Figure~\ref{fig:pos-trans} presents the corresponding $\sigma_{68}(\Delta r_{\perp})$ as a function of the same quantities.

In both figures, the resolution uncertainty grows at large distances to the detector wall, both along the particle direction and to the nearest wall. This is a statistical effect arising from the uniform vertex distribution and isotropic direction sampling used in the simulation: only a small fraction of events produce large wall distances, resulting in limited bin statistics and correspondingly wider confidence intervals at $\gtrsim\unit[8000]{cm}$ along the particle direction and $\gtrsim\unit[2500]{cm}$ to the nearest wall, consistent with the behavior observed in the lower panel of Fig.~\ref{fig:PID-signal-efficiency}.

\begin{figure*}[tp]
    \centering
    \includegraphics[width=1\textwidth]{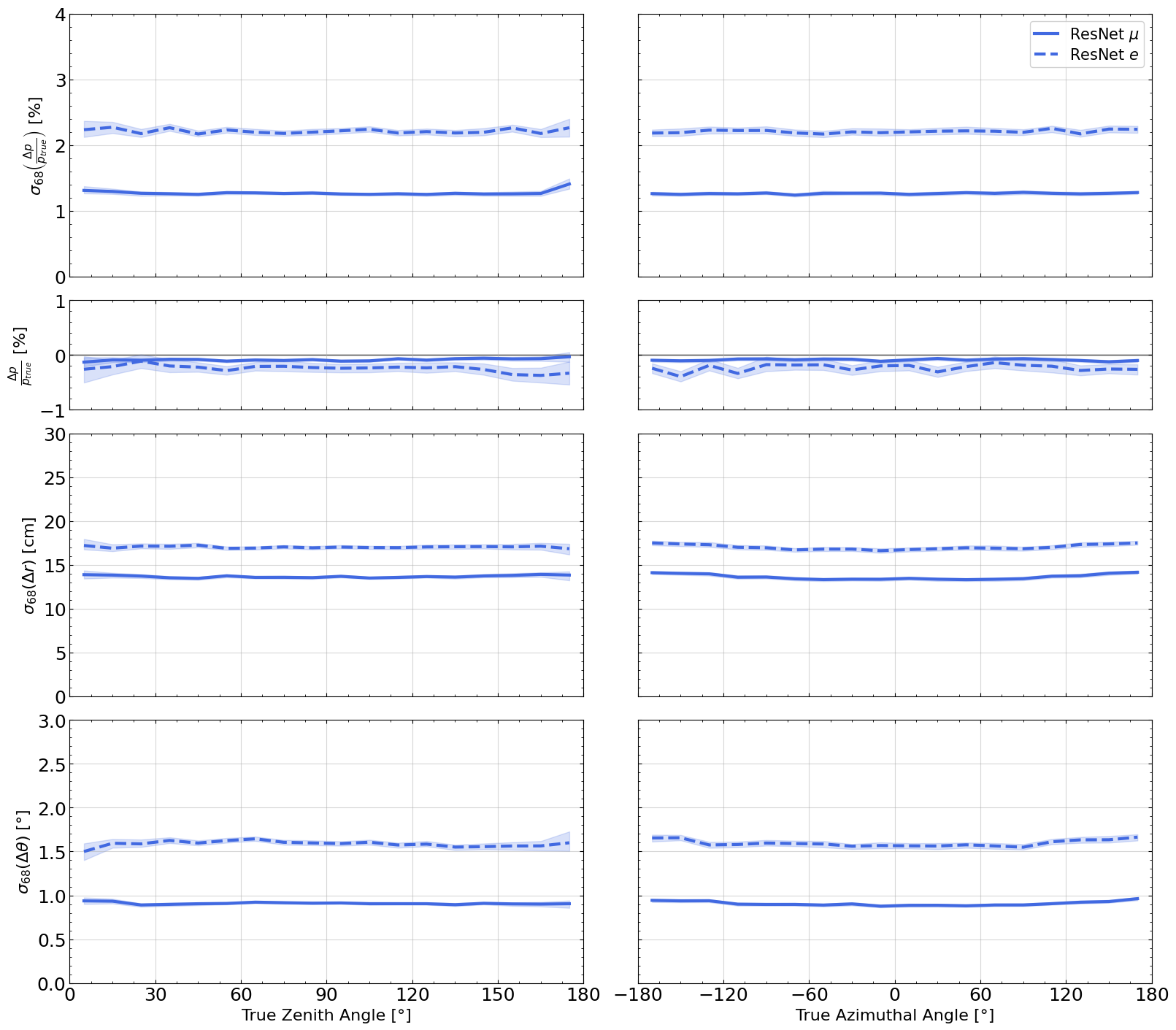}
    \caption{The same kinematic reconstruction performance metrics found in Section~\ref{sec:regression_results}, plotted against the true zenith angle (left column) and the true azimuthal angle (right column) of the event. A zenith angle of $0^{\circ}$ corresponds to the upward direction of the tank, $90^{\circ}$ to the horizontal direction (pointing towards the barrel wall), and $180^{\circ}$ to the downward direction. After the double-cover transformation, azimuthal angles of $-180^{\circ}$ and $180^{\circ}$ correspond to where the barrel is cut (i.e. the left and right edges of the image in Figure~\ref{fig:input-image}). Shaded bands are the $95\%$ bootstrap confidence intervals, though these are often narrower than the line width.}
    \label{fig:reg-az-zen}
\end{figure*}

Figure~\ref{fig:reg-az-zen} shows how the network performance does not degrade with zenith angle (that is, angle from the vertical axis of the tank) nor with azimuthal angle (i.e. the angle along the tank wall). We show the same kinematic reconstruction metrics found in Section~\ref{sec:regression_results} but plotted against true zenith angle (left column) and true azimuthal angle (right column). After the double-cover transformation, azimuthal angles of $-180^{\circ}$ and $180^{\circ}$ correspond to the left and right edges of the image in Figure~\ref{fig:input-image}, where the barrel is cut. No azimuthal or zenith dependence is observed, suggesting that the double-cover is sufficient to prevent any bias from the seam created by unfolding the tank.

\begin{figure}[!htbp]
    \centering
    \includegraphics[width=\linewidth]{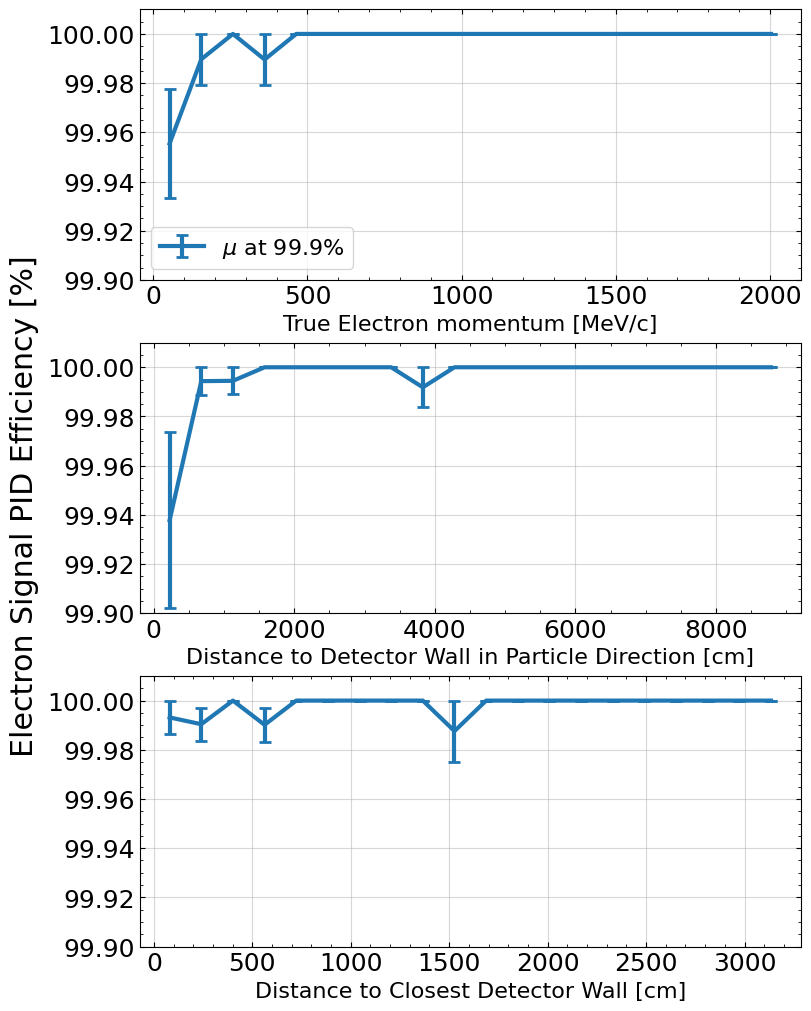}
    \caption{Same as Fig.~\ref{fig:PID-signal-efficiency}, but zoomed into the ResNet $e$-$\mu$ classification performance to allow for improved resolution. Error bars show the binomial standard error on the efficiency in each bin.}
    \label{fig:PID-signal-efficiency-zoomed}
\end{figure}

Figure~\ref{fig:PID-signal-efficiency-zoomed} reproduces the $e$-$\mu$ signal efficiency curve from Fig.~\ref{fig:PID-signal-efficiency}, but with a zoomed view of the ResNet classifier performance. This enhanced resolution highlights differences in discrimination power that are less apparent in the full-scale plot, particularly in the low background-acceptance regime relevant for high-purity electron selection.

\begin{figure}[!htbp]
    \centering
    \includegraphics[width=\linewidth]{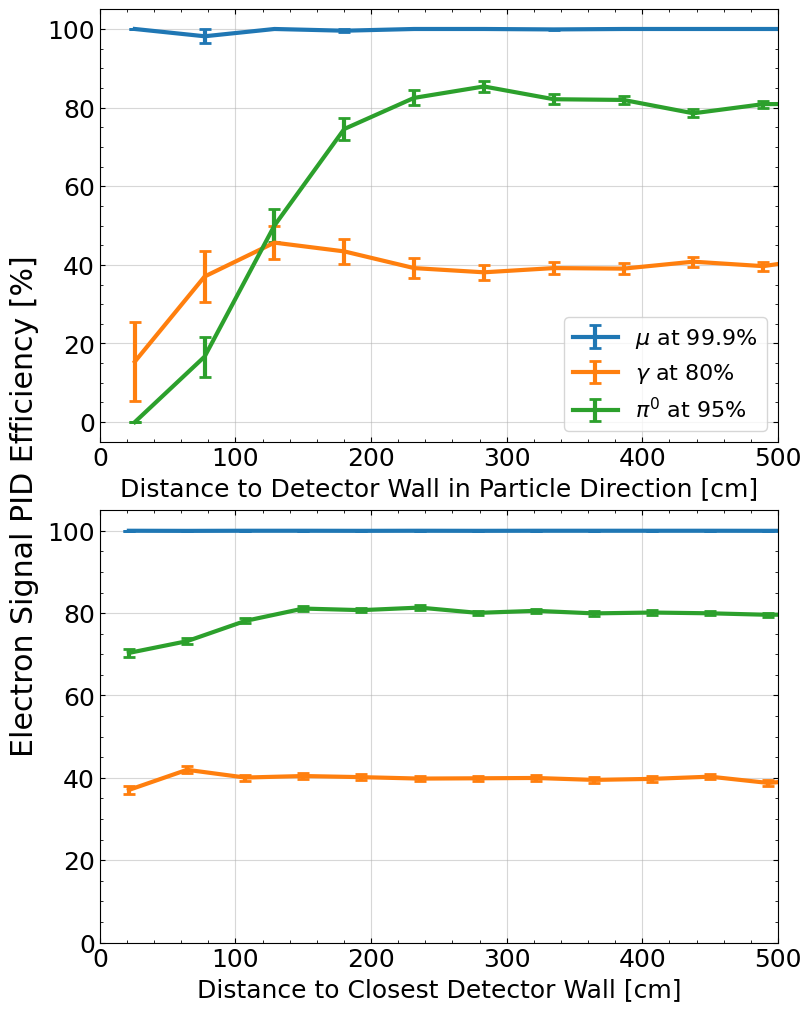}
    \caption{Same as Fig.~\ref{fig:PID-signal-efficiency}, but with the distance-to-wall axes restricted to $<\unit[500]{cm}$ to better resolve classifier performance for near-wall events: along the particle direction (top panel) and to the nearest detector wall (bottom panel). Error bars show the binomial standard error on the efficiency in each bin.}
    \label{fig:PID-signal-efficiency-small-dwall/towall}
\end{figure}

Figure~\ref{fig:PID-signal-efficiency-small-dwall/towall} reproduces the classifier efficiency profiles from Figure~\ref{fig:PID-signal-efficiency}, but restricts the distance axes to $\leq\unit[500]{cm}$ to better resolve near-wall behavior: along the particle direction (top panel) and to the nearest detector wall (bottom panel). The bin width is reduced accordingly. At very small distances to the wall along the particle direction, the classification performance degrades significantly, as the Cherenkov cone intersects the detector boundary at short range and most of the resulting light is collected by a single PMT or a small cluster of tubes, substantially reducing the available topological information. This behavior is expected.

\section{Effect of Muon Decay} \label{sec:decayeffect}
At lower momenta, muons are less relativistic and stop within the detector before their kinetic energy falls below the Cherenkov threshold. A stopped muon will subsequently decay into a Michel electron with a lifetime of $\tau_\mu = 2200\,\mathrm{ns}$. The probability of this decay occurring within the $950\,\mathrm{ns}$ post-trigger readout window is $P = 1 - e^{-950/\tau_\mu} \approx 35\%$. Since almost all muons below $\unit[250]{MeV/c}$ stop within the tank, and even higher-energy muons typically travel only $5$--$10\,\mathrm{m}$ before stopping (meaning only $\sim10\%$ exit the detector), a substantial fraction of events across all momentum bins will contain a Michel electron within the readout window. We verified this numerically in a sample of $10^{4}$ events with a uniform energy distribution, finding that $30$-$35\%$ of events in each momentum bin contain a decay electron within the window. If this decay electron contributes to the event image, it effectively produces a multi-ring topology that could confuse the reconstruction networks. Although time-clustering techniques should in principle help disentangle the prompt muon light from the delayed Michel-electron contribution, we investigate here how permitting decays affects reconstruction performance in practice.

\begin{figure}[!htbp]
\centering
\includegraphics[width=1\linewidth]{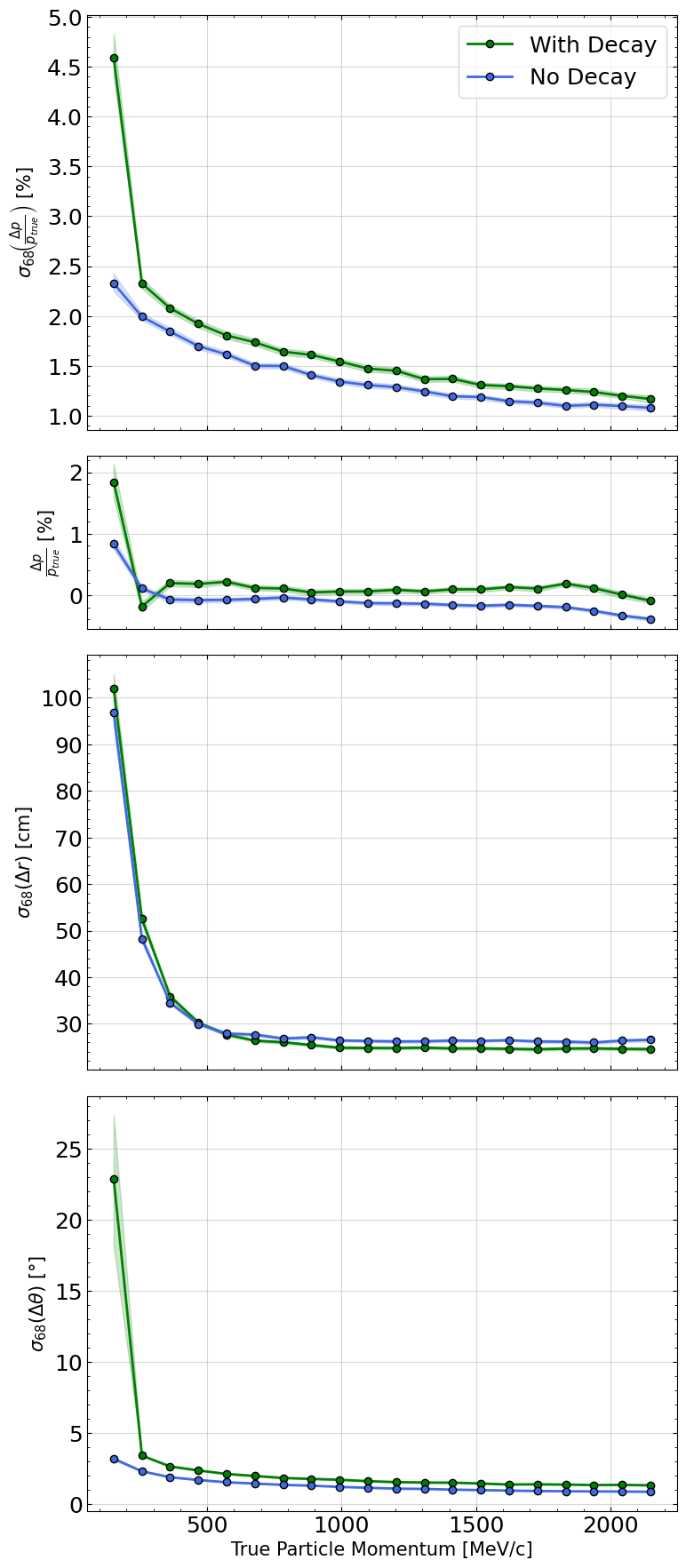}
\caption{Muon regression performance with and without decay. Plots in green correspond to networks trained on muon events where decay into an electron is permitted; baseline results (blue) from networks trained without muon decay are included for comparison. From top to bottom: momentum resolution (cf.\ Figure~\ref{fig:mom-reg}), momentum bias (cf.\ Figure~\ref{fig:mom-reg}), Euclidean vertex reconstruction error (cf.\ Figure~\ref{fig:pos-reg}), and direction reconstruction (cf.\ Figure~\ref{fig:dir-reg}). Shaded bands indicate $95\%$ bootstrap confidence intervals, though these are often narrower than the line width throughout.}
\label{fig:muondecay}
\end{figure}

To quantify this effect, we generate an additional dataset of $10$ million muon events in which decays are permitted and train a separate ensemble of regression networks for momentum, position, and direction reconstruction. Apart from allowing decay, the training procedure and event-selection cuts are identical to those described in Section~\ref{sec:architecture}. However, neither the decay-enabled nor the decay-disabled networks shown in Figure~\ref{fig:muondecay} apply the double-cover transformations or the hit-time invariance transformations used for the networks described in Section~\ref{sec:architecture}. This does not affect the comparison, since both ensembles described here lack these transformations, but it is worth noting that the decay-disabled results in Figure~\ref{fig:muondecay} will therefore differ slightly from the decay-disabled results shown in Section~\ref{sec:regression_results}.

Allowing muon decay has two primary consequences for reconstruction. First, the momentum regression degrades most significantly at low momenta ($\lesssim250\,\mathrm{MeV/c}$), exhibiting a $\sim2.3\%$ worse resolution and $\sim1\%$ worse bias. This occurs because at these energies the initial muon produces relatively few Cherenkov photons, so the Michel electron can dominate the total hit pattern. Even having been trained on decaying events, the network interprets the additional photons as increased muon energy, since at these low momenta the Michel electron dominates the charge pattern and the muon signal provides insufficient information to disentangle the two contributions. At higher momenta, the muon itself dominates the charge and timing distributions, and the effect of the Michel electron is correspondingly smaller: we observe only a modest $\sim0.2\%$ worsening in momentum resolution across the full range above $\unit[250]{MeV/c}$, with similarly minor degradation in direction reconstruction.

Second, direction reconstruction is substantially impacted at low momenta. For events with $\lesssim250\,\mathrm{MeV/c}$, the predicted direction can deviate by $\sim20^{\circ}$. In this regime, the muon ring is comparatively faint while the electron ring dominates, and because the decay electron is emitted isotropically relative to the muon's initial trajectory, the network often infers a direction closer to the decay electron rather than the parent muon.

Interestingly, position reconstruction benefits slightly from decay across the full momentum range. Because stopped muons travel only a short distance from the interaction vertex before decaying, the Michel electron is produced close to the primary vertex and provides additional localized light. This supplementary spatial information appears to aid the network in localizing the vertex more accurately.

Overall, the dominant effect of muon decay is confined to momenta below $\unit[250]{MeV/c}$, where the Michel electron can dominate the event image and significantly bias momentum and direction reconstruction. Above this threshold, the impact is modest. Future work would benefit from more sophisticated treatment of these low-energy edge cases, for instance through explicit hit-time clustering to isolate the prompt muon signal prior to reconstruction.

\section{Training Curves} \label{sec:training-curves}

\begin{figure*}[tp]
    \centering
    \includegraphics[width=0.9\textwidth]{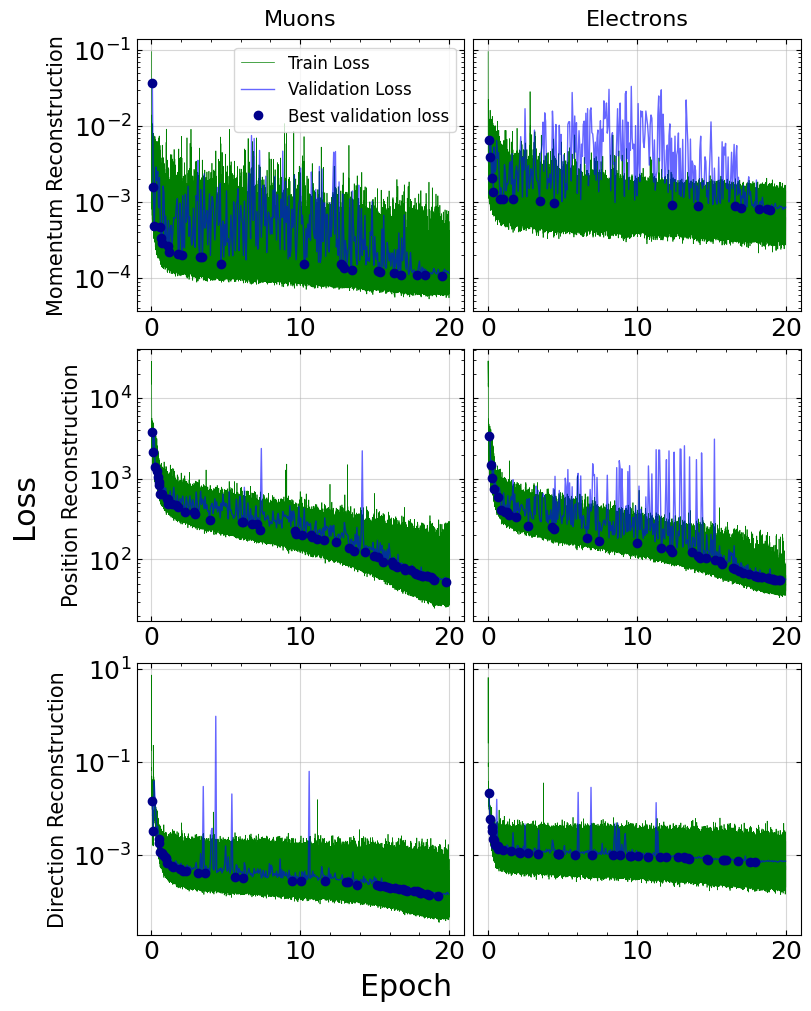}
    \caption{Loss curves for the six regression networks introduced in Section~\ref{sec:architecture} and evaluated in Section~\ref{sec:regression_results}. Panels on the left correspond to muon networks, and panels on the right to electron networks. From top to bottom: momentum, position, and direction regression. Training losses are shown in green and evaluated every iteration, while validation losses are shown in blue and evaluated every 2000 iterations. Blue markers denote the best validation loss achieved up to that point in training.}
    \label{fig:loss-curves}
\end{figure*}

Figure~\ref{fig:loss-curves} shows the training and validation loss curves for the regression networks described in Section~\ref{sec:regression_results}. The panels on the left correspond to muon networks, while those on the right correspond to electron networks.

Each plot displays the full 20-epoch training cycle. At regular intervals, both the training loss and validation loss are evaluated. Specifically, training proceeds for 2000 iterations with mini-batches of size 256 ($5.12\times10^{5}$ events), after which it is paused and the validation loss is computed using 690 mini-batches of 256 events ($1.77\times10^{5}$ events). The best validation loss up to that point is indicated with blue markers.

To mitigate overfitting and select the optimal model for each task, the network parameters are checkpointed whenever a new best validation loss is observed. After completing all 20 epochs, the parameters corresponding to the final best validation checkpoint are restored and used to evaluate the test set described in Section~\ref{sec:architecture}, yielding the results presented in Section~\ref{sec:regression_results}.

\section{WCSim Configuration for Hyper-Kamiokande Far Detector Simulations} \label{sec:wcsim-config}

The Monte Carlo samples used in this study were generated with version 1.12.19 of \textsc{WCSim} \cite{WCSim} using the 
\texttt{HyperK\_HybridmPMT\_WithOD\_Realistic} geometry, which includes the 50-cm inner-detector PMTs, the multi-PMT modules, and the outer-detector instrumentation.
For the purposes of this work, only the 50-cm inner-detector PMTs are used for reconstruction. The primary configuration settings are summarized below.

\paragraph*{Geometry and PMT response}
The full inner+outer detector geometry was enabled via the standard \textsc{WCSim} construction.
PMT quantum efficiency was applied using the \texttt{SensitiveDetector\_Only} method, and PMT collection efficiency was enabled. 
All other optical  properties, such as attenuation, scattering, and PMT timing response, followed the default Hyper-Kamiokande configuration distributed with \textsc{WCSim}.

\paragraph*{Digitization and trigger}
All PMT systems (50-cm inner-detector, multi-PMT, and outer-detector channels) used the \texttt{SKI} digitizer together with the \texttt{NDigits} trigger.
The digitizer used a deadtime of 560\,ns, a charge-time integration window of 200\,ns, timing precision of 0.1\,ns, and charge precision of 0.3\,photo-electrons.
Multiple digits per PMT per trigger were allowed. The \texttt{NDigits} trigger used a threshold of 25 digits within a 200\,ns window, with an automatic adjustment for noise occupancy.
For each trigger, hits from $-400$\,ns to $+950$\,ns around the trigger time were recorded, defining the 1.35\,$\mu$s readout window used throughout this work.
Only triggered events were written to file.

\paragraph*{Dark noise model}
Dark noise was added to the 50-cm inner-detector PMTs at a rate of 4.2\,kHz per tube using mode~1, corresponding to a time window of 4000\,ns centered on each hit.
This configuration approximates the expected dark-noise occupancy during the readout gate.
Dark noise for the 3-inch outer-detector PMTs and for the 3-inch PMTs within the multi-PMT modules was set to zero and therefore disabled.

\paragraph*{Physics configuration and event generation}
Events were generated using the GPS single-particle generator (\texttt{/mygen/generator gps}) with an isotropic angular distribution (\texttt{/gps/ang/type iso}).
Interaction vertices were drawn uniformly throughout the inner detector volume, modeled as a cylinder of radius 32.4\,m and half-height 32.9\,m (\texttt{/gps/pos/radius 3240 cm}, \texttt{/gps/pos/halfz 3290 cm}) and centered at the detector origin.
Separate macro files were used for each primary species, with the kinetic energy ranges chosen to span from the relevant Cherenkov threshold up to 2\,GeV above threshold where appropriate.
For the electron sample, \texttt{/gps/particle e-} was used together with a uniform energy distribution between 0.264\,MeV and 2000.264\,MeV (\texttt{/gps/ene/min 0.264 MeV}, \texttt{/gps/ene/max 2000.264 MeV}).
The lower bound corresponds to the electron kinetic Cherenkov threshold in water, so the kinetic energy range covers 0-2\,GeV above threshold.
For the muon sample, \texttt{/gps/particle mu-} was used with energies drawn uniformly between 54.6\,MeV and 2054.6\,MeV (\texttt{/gps/ene/min 54.6 MeV}, \texttt{/gps/ene/max 2054.6 MeV}), again corresponding to muon kinetic energies from the Cherenkov threshold up to 2\,GeV above threshold. 
For the gamma sample, \texttt{/gps/particle gamma} was used in combination with 
\texttt{/mygen/generator gamma-conversion}, which forces all photons to undergo pair production.
The photon energy was sampled uniformly between 1.55\,MeV and 2001.55\,MeV (\texttt{/gps/ene/min 1.55 MeV}, \texttt{/gps/ene/max 2001.55 MeV}), ensuring that the resulting $e^{+}e^{-}$ pair is always above the electron Cherenkov threshold and therefore produces visible Cherenkov light.
For the neutral-pion sample, \texttt{/gps/particle pi0} was used with kinetic energies drawn uniformly between 0\,MeV and 2000\,MeV (\texttt{/gps/ene/min 0 MeV}, \texttt{/gps/ene/max 2000 MeV}); the \(\pi^{0}\) mesons promptly decay into two 
photons, which then generate Cherenkov light through their electron-positron daughters when above threshold. 
In the electron, gamma, and \(\pi^{0}\) samples, muon decay and muon nuclear capture processes were left enabled as provided by the default \textsc{WCSim} physics list.
For the muon sample, however, both muon decay and \(\mu^{-}\) nuclear capture at rest were explicitly disabled to ensure clean single-muon topologies without Michel electrons or capture products. 
All hit-producing particle tracks were saved (\texttt{/Tracking/saveHitProducingTracks true}), and no optical-photon thinning was applied (\texttt{/Tracking/ fractionOpticalPhotonsToDraw 0.0}).

\section{Double-Cover Transformation} \label{app:double-cover}

Here we describe the double-cover transformation introduced in Section~\ref{sec:data}. The input image is first shifted horizontally by a quarter of its width. A $180^{\circ}$-rotated copy of each endcap is then inserted half a circumference away in azimuth, occupying pixels that were previously empty. Finally, the barrel is cut along its horizontal midline and each half is rotated by $180^{\circ}$ and appended above and below the image. The resulting $366\times144$ image is what is passed to the networks.

\begin{figure}[!htbp]
    \centering
    \includegraphics[width=0.99\linewidth]{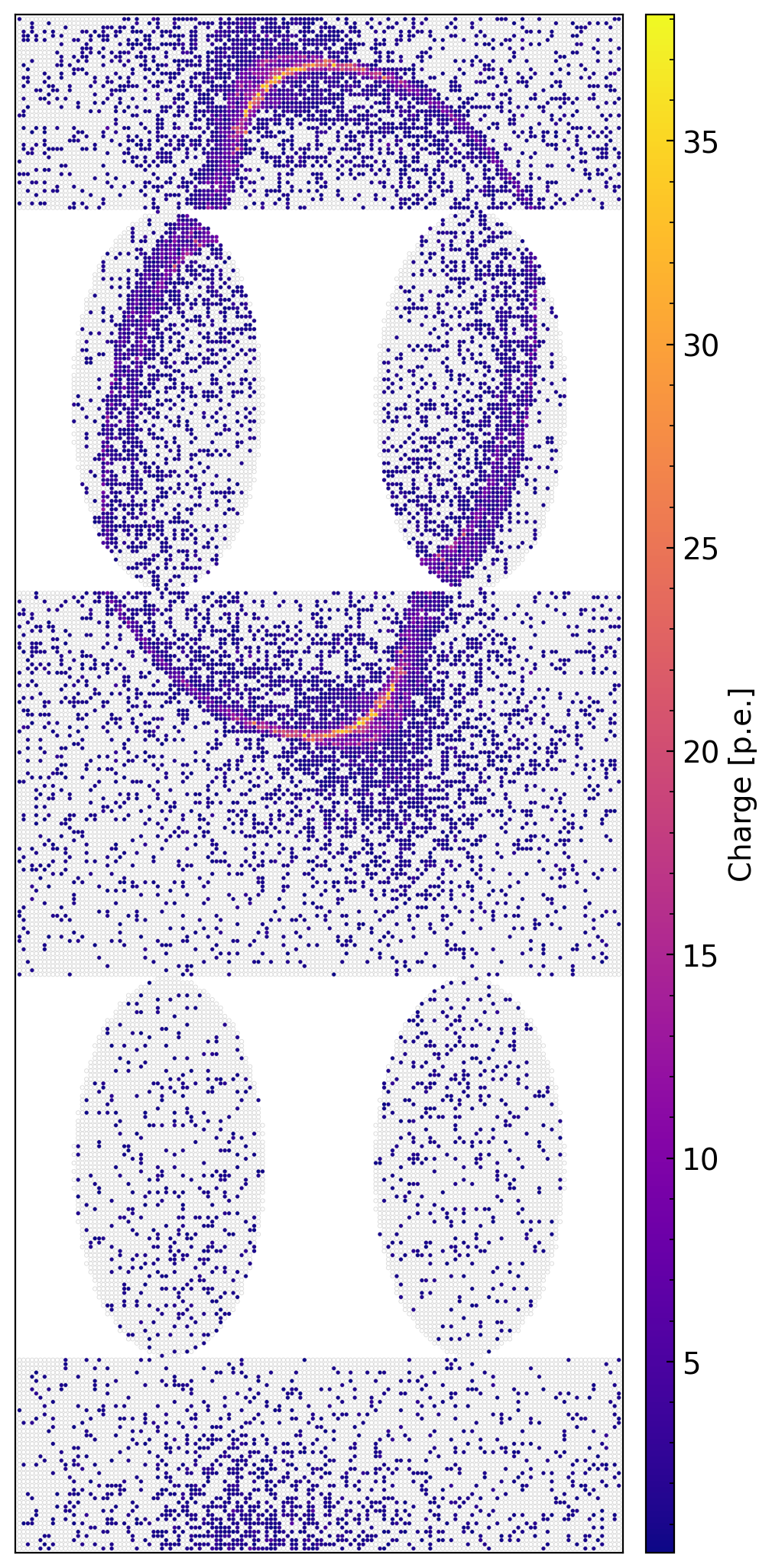}
    \caption{The same example event as Figure~\ref{fig:input-image}, after the double-cover transformation. Every PMT now appears twice, once near an image edge and once in the interior, so that a Cherenkov ring truncated at an edge in one view remains intact in the other.}
    \label{fig:double-cover-input-image}
\end{figure}

\begin{figure}[!htbp]
    \centering
    \includegraphics[width=1\linewidth]{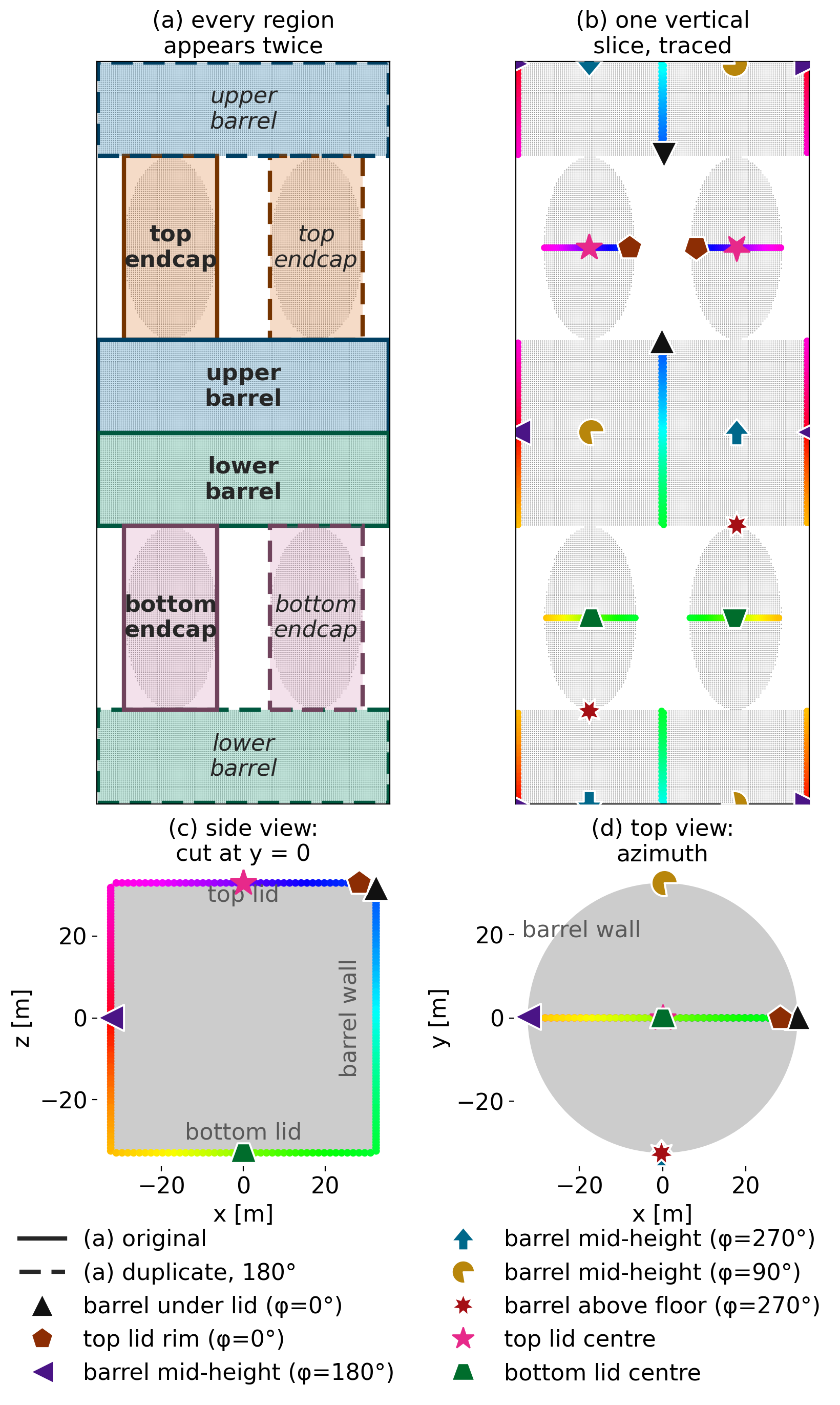}
    \caption{The double-cover representation of the Hyper-Kamiokande PMT image, in which the endcaps and barrel halves are each duplicated and rotated by $180^{\circ}$ to give a $366\times144$ image with cyclic boundaries. (a) Block decomposition; solid outlines are originals, dashed outlines the rotated duplicates. (b) Double-cover image where the rainbow line traces one continuous path around a hypothetical slice through the detector showing where a physically connected route ends up in the image. The markers indicate specific locations in the detector, each of which appears in exactly two places; markers in a duplicated block are drawn rotated by $180^{\circ}$, and those on an edge continue across the opposite edge, showing that the image is cyclic in both directions. (c, d) The same landmarks in a cut at $y = 0$ (i.e. seen from the side) and viewed from above.}
    \label{fig:double-cover-explainer}
\end{figure}

Every PMT therefore appears twice, with the barrel appearing both in the interior of the image and, inverted, at its upper and lower edges, so that a pattern truncated by the edge of the image in one copy lies away from the edge in the other. The rearrangement also reduces the fraction of the image that carries no information. Since the duplicated endcaps fill previously empty pixels, the fraction of pixels containing a PMT rises from $50\%$ to $75\%$. Figure~\ref{fig:double-cover-input-image} shows the same example event as Figure~\ref{fig:input-image} with the double-cover transformation applied, and Figure~\ref{fig:double-cover-explainer} illustrates how positions in the image map onto the detector.

\section{Absolute Hit-Time Invariance} \label{app:hit-time-invariance}
Invariance to a global time translation of all hit times is enforced by constraining the weights of the first convolutional layer of the network. The output of the convolutional layer, $y$ for channel $k$ and with spatial dimensions over $i$ and $j$, is given by
\begin{equation}
    y(k,i,j) = \sum_m \sum_n \sum_c w(k,c,m,n)\cdot x(c,i+m,j+n)+b(k),
\end{equation}
where $x$ is the input image with channels $c$, $b$ represents the learned biases, $w$ represents the learned weights of the convolutional kernels with spatial dimensions over $m$ and $n$. If the subset of input channels $T$ (in our case, only one channel corresponding to hit times) is shifted by a global constant $A$ (corresponding to the arbitrary trigger time of the event) then the modified output of the convolutional layer becomes
\begin{equation}
\begin{split}
    y'(k,i,j) &= \sum_{m,n}\sum_{c\in T} w(k,c,m,n)\,\bigl(x(c,i+m,j+n)+A\bigr) \\
    &\quad + \sum_{m,n}\sum_{c\notin T} w(k,c,m,n)\,x(c,i+m,j+n) + b(k) \\
    &= y(k,i,j) + A\sum_{m,n}\sum_{c\in T}w(k,c,m,n).
\end{split}
\end{equation}
When the sum of the weights $\sum_m\sum_n\sum_{c\in T}w(k,c,m,n)$ is enforced to equal zero, for each output channel $k$, then $y'(k,i,j) = y(k,i,j)$ and every value in the layer's output is invariant under any value of $A$. Because every subsequent layer acts solely on the output of the first, the invariance propagates through the network. In practice, this constraint is enforced by shifting these weights by subtracting their own mean, for each output channel, before applying the convolution, thereby ensuring the sum equals zero. The weights acting on the charge channels are left unconstrained, since the absolute scales of the charge values are physically meaningful.

The invariance holds when all input values $x$, for the time channels $c\in T$, are shifted by the same global offset $A$. This requires that the values used in the time channel of the input image pixels corresponding to unhit PMTs, or pixels that have no PMT, are assigned some value that translates along with any time-translation of hit times. For this reason, the median hit time is used for the values of the time channel of these pixels, rather than a fixed constant, while the charge channel is set to zero, as described in Section~\ref{sec:data}. The same approach is used for pixels added during padding of the image in the first convolution layer, rather than the standard zero-padding.
\end{document}